\documentclass[aps,letterpaper,superscriptaddress,pra,showpacs,amsmath,reprint,floats]{revtex4-1}

\usepackage[dvips]{graphicx}
\usepackage{amssymb,amsfonts,amsmath}
\usepackage{subfigure}
\usepackage{pdfpages}
\usepackage{braket}
\usepackage{bm}
\usepackage[utf8]{inputenc}
\usepackage[T1]{fontenc}
\usepackage{lmodern} 
\usepackage{amssymb,amsfonts,amsmath}
\usepackage{mathtools}
\usepackage{alltt}
\usepackage{mathrsfs}

\begin{document}

\begin{alltt}
\textcolor{blue}{\textbf{\large{Accepted for Publication in Physical Review A}}}
\end{alltt}

\title{An N-atom Collective State Atomic Interferometer with Ultra-High Compton Frequency and Ultra-Short de Broglie Wavelength, with root-N Reduction in Fringe Width}

\author{Resham Sarkar}
\email{rsarkar@u.northwestern.edu}
\affiliation{Department of Physics and Astronomy, Northwestern University, 2145 Sheridan Road, Evanston, IL 60208, USA}

\author{May E. Kim}
\affiliation{Department of Physics and Astronomy, Northwestern University, 2145 Sheridan Road, Evanston, IL 60208, USA}

\author{Renpeng Fang}
\affiliation{Department of Physics and Astronomy, Northwestern University, 2145 Sheridan Road, Evanston, IL 60208, USA}

\author{Selim M. Shahriar}
\affiliation{Department of Physics and Astronomy, Northwestern University, 2145 Sheridan Road, Evanston, IL 60208, USA}
\affiliation{Department of EECS, Northwestern University, 2145 Sheridan Road, Evanston, IL 60208, USA}

\begin{abstract}
We describe a collective state atomic interferometer (COSAIN) with the signal fringe as a function of phase-difference or rotation narrowed by $\sqrt{N}$ compared to a conventional interferometer - $N$ being the number of atoms - without entanglement. This effect arises from the interferences among collective states, and is a manifestation of interference at a Compton frequency of ten nonillion Hz, or a de Broglie wavelength of $4.5$ femtometer, for $N=10^6$ and $v = 1$ m/s.  The population of the collective state of interest is detected by a null measurement scheme, in which an event corresponding to detection of zero photons corresponds to the system being in that particular collective state. The signal is detected by collecting fluorescence through stimulated Raman scattering of Stokes photons, which are emitted predominantly against the direction of the probe beam, for a high enough resonant optical density. The sensitivity of the ideal COSAIN is found to be given by the standard quantum limit.  However, when detection efficiency and collection efficiency are taken into account, the detection scheme of the COSAIN increases the quantum efficiency of detection significantly in comparison to a typical conventional Raman atomic interferometer employing fluorescence detection, yielding a net improvement in stability by as much as a factor of $10$. We discuss how the inhomogeneities arising from the non-uniformity in experimental parameters affect the COSAIN signal.  We also describe an alternate experimental scheme to enhance resonant optical density in a COSAIN by using cross-linearly polarized counter-propagating Raman beams.

\end{abstract}


\maketitle

\section{\label{sec:intro}Introduction}
Matter wave interferometry is a potent technology in metrology. Atom interferometers have been demonstrated as gyroscopes and accelerometers~\cite{Kasevich,Canuel}, gravity gradiometers~\cite{Chu,Snadden}, matter-wave clocks~\cite{Lan} and may lead to a more accurate measurement of the fine structure constant~\cite{Biraben,Cadoret}. They also form testbeds for measuring Newton's gravitational constant~\cite{Fixler}, gravitational red-shift~\cite{Muller} and for testing universality of free fall~\cite{Geiger}.

\begin{figure}[!h]
\includegraphics[scale=0.5]{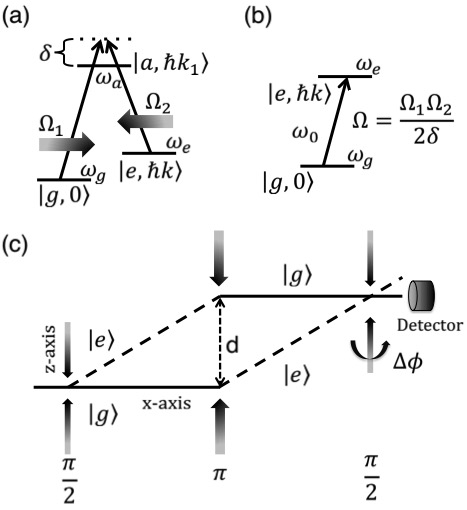}
\caption{\label{Fig_1}(a) A three level atom. (b) An equivalent reduced two-level atom model. (c) A CRAIN produced via $\pi/2-\pi-\pi/2$ sequence of excitation.}
\end{figure}

The building block of a Conventional Raman Atom Interferometer (CRAIN) is a three level atom, with two metastable states, $\ket{g,p_z\!=\!0}\equiv\ket{g,0}$ and $\ket{e,p_z\!=\!\hbar(k_1\!+\!k_2)}\equiv\ket{e,\hbar{k}}$ and an excited state $\ket{a,p_z\!=\!\hbar{k_1}}\equiv\ket{a,\hbar{k_1}}$ coupled by two counter propagating beams, with a single photon detuning $\delta$ (Fig.~\ref{Fig_1}(a)). One of the beams, with Rabi frequency $\Omega_1$, couples $\ket{g,0}$ to $\ket{a,\hbar{k_1}}$, while the other beam, with Rabi frequency $\Omega_2$, couples $\ket{a,\hbar{k_1}}$ to $\ket{e,\hbar{k}}$. For $\delta\!\gg\!\Omega_1,\Omega_2$, the interaction can be described as an effective two level system excited by an effective traveling wave with a momentum $\hbar{k}\!=\!\hbar(k_1\!+\!k_2)$, with a Rabi frequency $\Omega\!=\!\Omega_1\Omega_2/2\delta$ (Fig.~\ref{Fig_1}(b))~\cite{Shahriar}. We assume that $\delta\!\gg\!\Gamma$, where $\Gamma$ is the decay rate of $\ket{a}$, so that the effect of $\Gamma$ can be neglected. Under a sequence of $\pi/2-\pi-\pi/2$ pulses (Fig.~\ref{Fig_1}(c)), the wavepacket first separates into two components, then gets redirected and finally recombined to produce an interference which is sensitive to any phase-difference, $\Delta\phi$ between the two paths. The amplitude of $\ket{g}$ at the end varies as $\cos^{2}(\Delta\phi/2)$~\cite{KasevichChu,Borde}.

A CRAIN of this type can be realized by employing an atomic beam with a continuous flux, or by employing pulses of atoms pushed out periodically from a magneto-optic-trap (MOT). The behavior of the CRAIN is essentially the same in both modes if the number of atoms interrogated in a given time window is the same. However, as we will describe later, the collective state atomic interferometer (COSAIN) must operate in the latter (pulsed) mode. Thus, for proper comparison we will assume, in the rest of the paper, that the CRAIN is operated in the pulsed mode.

The phase-difference induced due to rotation at the rate of $\Omega_G$ along an axis normal to the area, $\Theta$ of the interferometer is given by, $\Delta\phi\!=\!4\pi\Theta{m}\Omega_G/h$, $m$ being the atomic mass~\cite{Borde,Riehle}. This expression can be derived by two different methods. In the first method, the path-difference of the two counter-propagating waves is multiplied by $2\pi/\lambda_{dB}$, where $\lambda_{dB}$ is the de Broglie wavelength, to get the phase-difference. The second method invokes the relativistic addition of velocities to find the time lag, $\Delta{T}=2\Theta\Omega_G/c^2$ in the arrival of the two branches of the wave, $c$ being the speed of light. $\Delta\phi$ is then the product of $\Delta{T}$ and the wave frequency. For the CRAIN, this frequency is the Compton-frequency of the atom, $\omega_C\!=\!\gamma\!mc^2/\hbar\!\approx\!mc^2/\hbar$, where the relativistic time-dilation factor, $\gamma$, is close to unity for non-relativistic velocities. These approaches are equivalent due to the fact that $\lambda_{dB}$ is the laboratory-frame manifestation of the $\omega_C$ induced phase variation in the rest-frame of the atom~\cite{Dowling,Malykin,Broglie,Lan}. To explain this without loss of generality, let us consider the direction of the velocity of the particle as  $\mathbf{\hat{x}}$. For non-relativistic velocities, mixing between the spinors can be ignored, and the phase factor of a positive energy spinor, in the rest frame of the particle, is given simply as $\exp(-i\phi)$, where $\phi=\omega_C\tau$ with $\tau$ being the proper time. The phase,  $\phi$, is a Lorentz invariant parameter, and can in general be written as a contraction between the position four vector $x^{\mu}$ and momentum four vector $\hbar k^{\mu}: \phi = k_{\mu}x^{\mu}$. In the rest frame of the particle, the position four-vector is $x^{\mu}=\{c\tau,0,0,0\}$ and the momentum four vector is $\hbar k^{\mu}=\hbar\{\omega_C/c,0,0,0\}$. In the laboratory frame, the position four vector is, by definition, $x^{\mu}=\{ct,0,0,0\}$, and application of Lorentz transform shows that the momentum four vector is $\hbar k^{\mu}=\hbar\{\omega_C'/c,k_{dB},0,0\}$, where $k_{dB}=\gamma mV/\hbar=2\pi/\lambda_{dB}$ and the phase factor becomes $\phi=\omega_C't-k_{dB}x$.  Again, in the non-relativistic limit, $\gamma\approx 1$ and we get $\lambda_{dB}\approx 2\pi\hbar/mV$.  Thus, the de Broglie wavelength is simply the laboratory frame manifestation of the phase variation in the rest frame due to the Compton frequency.

The dependence of $\Delta\phi$ on $\omega_C$ has motivated matter wave interferometry with large molecules. To date, the largest molecule used has a mass of $\sim10000$ atomic-mass-unit~\cite{Eibenberger}, corresponding to the mass of $\sim75\;^{133}$Cs atoms. These interferometers, based on the Talbot effect, are not suited for rotation sensing. Furthermore, for interferometry with much larger particles it would be necessary to use gratings with spacings too small to be realized with existing technologies. Additionally, effects such as van der Waals interaction would become dominant for such gratings. Here, we propose an experiment that would reveal evidence of matter wave interference where a collection of $N$ non-interacting, unentangled atoms acts as a \textit{single} particle. For $^{87}$Rb and $N=10^6$, $\omega_C$ is $\sim$ ten nonillion Hz, and $\lambda_{dB}$ is $\sim$ $4.5$ femtometer at a velocity of $1$ m/s. Furthermore, it can improve the phase measurement ability by a factor of as much as $10$. This type of matter wave interferometry may also open up new opportunities for sensitive measurement of gravitational redshift~\cite{Muller} or matter wave clocks~\cite{Lan}. It may also serve as a testbed for macroscopic quantum decoherence due to gravitational redshift~\cite{Pikovski}.

Consider an assembly of $N$ identical noninteracting atoms, subjected to the $\pi/2-\pi-\pi/2$ sequence. If we imagine a situation where the ground state, $\ket{E_0}\!\equiv\!\ket{g_1,g_2,\ldots,g_N}$ is coupled, directly and only, to the state where all the atoms are in the excited state, $\ket{E_N}\!\equiv\!\ket{e_1,e_2,\ldots,e_N}$, the resulting ensemble interferometer would experience a phase-difference, $\Delta\phi_{EI}\!=\!N\Delta\phi$. However, existing technology does not enable such an excitation. Even if one were to use a pure Fock state of $N'>N$ photons, the ensemble would evolve into a superposition of $(N+1)$ symmetric collective states $\ket{E_n}\ket{N'-n}$, where $\ket{N'-n}$ is a state of the field with $(N'-n)$ photons, and $\ket{E_n}\!=\!J(N,n)^{-1/2}\sum_{k\!=\!1}^{J(N,n)}P_k\ket{g^{\otimes(N\!-\!n)}e^{\otimes\!n}}$, where $J(N,n)\!\equiv\!{N \choose n},\!P_k$ is the permutation operator, and $n=0,1,2,\ldots,N$~\cite{HumeWineland}. Since a laser is a superposition of many Fock states, the evolution of this system under laser excitation would produce a seemingly intractable superposition of these collective states. Modeling the laser field as a semi-classical one also does not simplify the picture much~\cite{Dicke,Arecchi,Scully,CollectiveDescription}. However, we show here that, by measuring the quantum state of a single collective state, it is possible to determine the effect of the interference among all the collective states, and describe how such a measurement can be done. Choosing this collective state to be one of the two extremal states (i.e., $\ket{E_0}$ or $\ket{E_N}$) also makes it possible to calculate this signal easily, since the state of the whole system can be described as the tensor product of individual atomic states. We show that the fringe width is reduced by a factor of $\sqrt{N}$, without using entanglement. For the current state of the art, the value of $N$ can easily exceed $10^6$, so that a reduction of fringe width by a factor of more than $10^3$ is feasible. We also show that the phase fluctuation of the COSAIN can be significantly smaller, by as much as a factor of $10$, than that for a conventional interferometer employing the same transition and the same atomic flux. The extremely narrow resonances produced in the COSAIN may also help advance the field of spin squeezing~\cite{Ueda,Polzik,Kuzmich,SpinSqueezeRev}, which in turn is useful for approaching the Heisenberg limit in precision metrology. Recently, we have also proposed a collective state atomic clock, which employs the principle of collective excitation of atomic ensemble, and exhibits a similar narrowing in signal fringe~\cite{COSAC}.

In this paper, we discuss the various aspects of the collective state atomic interferometer. The rest of the paper is arranged in the following way: in Sec.~\ref{sec:description}, we describe the theory of the working principle of a COSAIN. We also describe the physical phenomenon behind the narrowing of the signal fringes. Sec.~\ref{sec:inhomogeneities} gives an account of the various parameter inhomogeneities that affect the signal amplitude and width. Sec.~\ref{sec:experiment} details the description of the COSAIN experiment, also including a discussion of the role of the optical density of the ensemble. We also propose an alternate experimental scheme to achieve a higher value of effective optical density in Sec.~\ref{sec:alternate}. In Sec.~\ref{sec:comparison}, we analyze the performance of the COSAIN as compared to that of the CRAIN. We consider the effect of quantum and classical noise, detector efficiency, and collection efficiency.

\section{\label{sec:description}Description of the COSAIN}

Consider an ensemble of $N$ noninteracting atoms of the kind described above~\cite{Dicke}, with the $i$-th atom in its ground state, $\ket{g_i}$. The ensemble is assumed to be initially situated at $(x\!=\!0, \!z\!=\!0)$ and traveling along the $\mathbf{x}$-direction with a velocity $v$. The ensemble undergoes the same $\pi/2-\pi-\pi/2$ sequence as described for the CRAIN. Assuming resonant excitation, the Hamiltonian of the $i$-th atom after the rotating-wave transformation is $H_i\!=\!\Omega_i\ket{g_i}\bra{e_i}/2\!+\!c.c.$~\cite{CollectiveDescription}, where $\Omega_i$ is the Rabi frequency of the $i$-th atom. Here, a phase transformation on the Hamiltonian has also been applied to render $\Omega_i$ real. For the sake of simplicity and brevity, we consider only the case where the intensity profile of the beams are rectangular, so that $\Omega_i\!=\!\Omega$. In a real experiment, the Rabi-frequency of each atom depends on its position relative to the Gaussian distribution of the beam intensity profile. Due to the non-zero temperature of the trapped atoms, they also experience Doppler shift arising from thermal motion. A detailed description of the effect of these inhomogeneities on the COSAIN signal is presented in Sec.~\ref{sec:inhomogeneities}.

A $\pi/2$-pulse of duration $\tau$ is applied to the ensemble at $t=0$, following which each atom is in state $\ket{\psi_i(\tau)}\!=\!(\ket{g_i}\!-i\ket{e_i})/\sqrt{2}$. After the first dark-zone of duration of $T_d$, the component of the atom in state $\ket{e_i}$ drifts to $(x=vT_d,z=\hbar{k}T_d/m)$. The state $\ket{g_i}$ continues along the \textbf{x}-direction. We label the trajectories taken by $\ket{g_i}$ and $\ket{e_i}$, $A$ and $B$ respectively. The state of an atom at $t\!=\!\tau\!+\!T_d$ is $\ket{\psi_i(\tau\!+\!T_d)}=\ket{\psi_i(\tau\!+\!T_d)}_A\!+\!\ket{\psi_i(\tau\!+\!T_d)}_B$, where $\ket{\psi_i(\tau\!+\!T_d)}_A\!=\!\ket{g_i}/\sqrt{2}$ and $\ket{\psi_i(\tau\!+\!T_d)}_B\!=\!-i\ket{e_i}/\sqrt{2}$. At the end of this zone, a $\pi$-pulse causes the state $\ket{g_i}$ to evolve into $\ket{e_i}$ and vice-versa. The state at the end of this pulse is $\ket{\psi_i(3\tau\!+\!T_d)}=\ket{\psi_i(3\tau\!+\!T_d)}_A\!+\!\ket{\psi_i(3\tau\!+\!T_d)}_B$, such that $\ket{\psi_i(3\tau\!+\!T_d)}_A\!=\!-i\ket{e_i}/\sqrt{2}$ and $\ket{\psi_i(3\tau\!+\!T_d)}_B\!=\!-\ket{g_i}/\sqrt{2}$. Following the second dark zone of duration $T_d$, the two trajectories converge, as shown in Fig.~\ref{Fig_1}(c), and $\ket{\psi_i(3\tau\!+\!2T_d)}=\ket{\psi_i(3\tau\!+\!T_d)}$. At $t=3\tau+2T_d$, a third pulse of duration $\tau$ is applied to the atoms. If a phase-difference of $\Delta\phi$ is introduced between the paths, the state of the atom at the end of the last $\pi/2$-pulse is $\ket{\psi_i(4\tau\!+\!2T_d)}\!=\!\ket{\psi_i(4\tau\!+\!2T_d)}_A\!+\!\ket{\psi_i(4\tau\!+\!2T_d)}_B$, where $\ket{\psi_i(4\tau\!+\!2T_d)}_A\!=\!-i(-i\exp(-i\Delta\phi)\ket{g_i}\!+\!\ket{e_i})/2$ and $\ket{\psi_i(4\tau\!+\!2T_d)}_B\!=\!-\!(\ket{g_i}\!-\!i\exp(i\Delta\phi)\ket{e_i})/2$. This phase-difference can occur, for example, due to a rotation of the entire system about the \textbf{y}-direction.

The final fringe pattern is the result of the interference of the states from the two trajectories. This is observed by measuring the probability of finding the atom in either of the two states. The signal as a measure of the amplitude of $\ket{g}$, is therefore, $S_{CRAIN}\!=\!|(1\!+\!\exp(-i\Delta\phi))/2|^2\!=\!\cos^2(\Delta\phi/2)$. We note now that the state $\ket{\Psi}$ of the ensemble is the direct product of its constituent atoms: $\ket{\Psi}\!=\!\prod_{i=1}^N\ket{\psi_i}$~\cite{Arecchi,CollectiveDescription}. The signal of the COSAIN is a measurement of any of the arising collective states. We choose to measure the state $\ket{E_0}$, so that the resulting signal is the probability of finding all the atoms of the ensemble simultaneously in $\ket{g}$. This choice of state will be explained later on when we discuss the detection system of the COSAIN. The signal of the COSAIN is thus the product of the signals from the constituent atoms, $S_{COSAIN}\!=\!\prod_{i=1}^N\!S_{CRAIN}\!=\!\cos^{2N}(\Delta\phi/2)$. The fringe linewidth as a function of $\Delta\phi$ decreases with increasing $N$. We define this linewidth as the full width at half maximum (FWHM) of the signal fringe, $\varrho(N)\!=\!2\cos^{-1}(2^{-1/2N})$. We have verified that $\varrho(1)/\varrho(N)\!\approx\!\sqrt{N}$.

\subsection{\label{sec:narrowing}Physical Interpretation of Fringe Narrowing}

The narrowing of the signal fringes in a COSAIN can be understood by considering the physical properties of the collective excitations. If the ensemble in the ground state interacts with a single photon of momentum $\hbar{k}$, it will oscillate between $\ket{E_0,0}\!\leftrightarrow\!\ket{E_1,\hbar{k}}$. Consequently, it will exhibit collective behavior such that its center of mass recoils with a velocity in the \textbf{z}-direction equal to $\hbar{k}/Nm$. Thus, this ensemble can be viewed as a single entity with a mass of $Nm$, and a Compton frequency,  $\omega_C$ that is $N$ times that of a single atom, despite no interaction between the atoms. Conversely, the ensemble can be viewed as having a $\lambda_{dB}$ of $h/Nmv$ that is $N$ times lower than that of a single atom, where $v$ is the magnitude of its total velocity (e.g., a constant velocity in the \textbf{x}-direction that is much larger than the velocity in the \textbf{z}-direction due to the recoil). In the ideal case of uniform Rabi-frequencies and no Doppler shift related detunings, the first $\pi/2$-pulse splits the ensemble into a superposition of $N+1$ symmetric collective states (we have shown the corresponding interpretation of the other, more general cases in ref.~\citenum{CollectiveDescription}). The state $\ket{E_n}$ receives a recoil of $n\hbar{k}$ due to the first $\pi/2$-pulse and is deflected in the \textbf{z}-direction by $n\hbar{k}T_d/Nm$ by the end of the first dark zone, making an angle $\theta_n\!=\!\tan^{-1}(n\hbar k/Nmv)$ with the \textbf{x}-axis. We label the path taken by this state as Path-$n$. The subsequent $\pi$-pulse causes $\ket{E_n}$ to evolve to $\ket{E_{N-n}}$. This results in the deflection of the trajectory of the states so that all the $N+1$ trajectories converge by the end of the second dark-zone. The third pulse causes each of the $N+1$ states to split further. The resulting COSAIN is, thus, $J(N+1,2)$ collective interferometers operating simultaneously. Of these, there are $x$ interferometers of area $(N-x+1)\Theta/N$, producing signal fringe amplitudes equaling $\cos^2((N-x+1)\Delta\phi/2)$, where $x$ assumes values $1,2,\ldots,N$. The interference between these cosinusoidal fringes result in the narrowing of the total fringe width. In what follows, we illustrate the physical mechanism behind this narrowing by considering first the role of Compton frequency in a CRAIN. We then extend this analysis to an ensemble of $N$ atoms to describe the phenomenon of narrowing in the COSAIN.

\begin{figure}[h]
\centering
\includegraphics[scale=0.5]{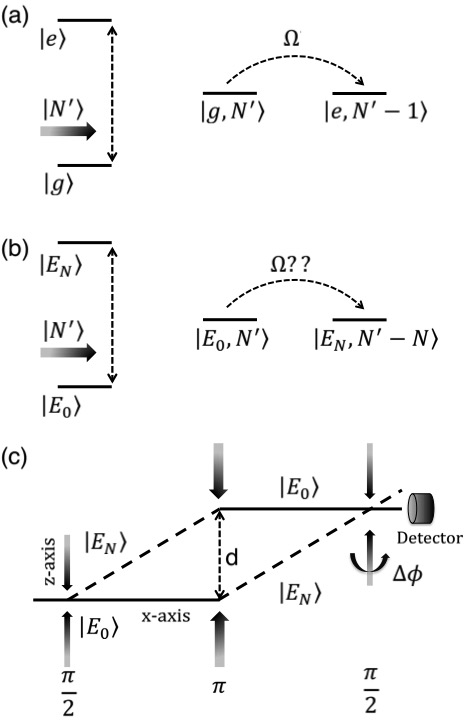}
\caption{\label{Fig_2}(a) Single atom coupled to an $N'$-photon state, (b)$N$-atom ensemble coupled to an $N'$-photon state, (c) Ensemble interferometer formed by splitting and recombining of $\ket{E_0}$ and $\ket{E_N}$.}
\end{figure}

We consider the product state of the atom and a Fock state with $N'$ photons denoted by $\ket{N'}$ or with $N'-1$ photons denoted by $\ket{N'-1}$. Thus, at $t=0$, the atom photon system is assumed to be in the state $\ket{g}\ket{N'}\equiv\ket{g,N'}$. The atom-field interaction couples it to the state $\ket{e}\ket{N'-1}\equiv\ket{e,N'-1}$, as illustrated in Fig.~\ref{Fig_2}(a). We assume that the photon energy, $\hbar\omega$, exactly matches the energy difference between the atomic internal states $\ket{e}$ and $\ket{g}$. We define the dressed frequency of the atom-photon system as $\omega_{PA}$, which is a constant, for all possible states of the system. If we define $\omega_{C,e}=m_ec^2/\hbar$ as the Compton frequency of the excited atom, where $m_e=m_g+\hbar\omega/c^2$ is the rest mass of the excited atom, and $m_g=m$ is the rest mass of the atom in the ground state, then we have $\omega_{PA}=m_ec^2/\hbar+(N'-1)\omega=m_gc^2/\hbar+N'\omega$. The Compton frequency of the atom in the ground state is $\omega_{C,g}=m_gc^2/\hbar$. The effect of temporal phase accumulation on the system during an interval $\Delta t$, if the system is in an arbitrary superposition of $\ket{g}$ and $\ket{e}$, i.e. $c_g\ket{g}+c_e\ket{e}$ at the start of the interval, will be $\exp(-i\omega_{PA}\Delta t)(c_g\ket{g,N'}+c_e\ket{e,N'-1})$. Thus, after the first $\pi/2$ pulse of a time duration, $\tau$, the quantum state of the system is $\exp(\omega_{PA}\tau)(\ket{g,N'}_A-i\ket{e,N'-1}_B)/\sqrt{2}$, where the subscripts $A$ and $B$ indicate the lower and upper trajectory of the interferometer, respectively. This is followed by a dark zone of duration $T_d$ at the end of which the quantum state of the system is $\exp(-i\omega_{PA}(\tau+T_d))(\ket{g,N'}_A-i\ket{e,N'-1}_B)/\sqrt{2}$. A $\pi$-pulse is applied at the end of the first dark zone, and therefore, at $t=3\tau+T_d$, the quantum state of the system is $\ket{\psi(3\tau+T_d)}=\ket{\psi(3\tau+T_d)}_A+\ket{\psi(3\tau+T_d)}_B$, where $\ket{\psi(3\tau+T_d)}_A=-i\exp(-i\omega_{PA}(3\tau+T_d))\ket{e,N'-1}/\sqrt{2}$ and $\ket{\psi(3\tau+T_d)}_B=-\exp(-i\omega_{PA}(3\tau+T_d))\ket{g,N'}/\sqrt{2}$. At this point the second dark zone begins, at the end of which the state of the system can be written as $\ket{\psi(3\tau+2T_d)}=\ket{\psi(3\tau+2T_d)}_A+\ket{\psi(3\tau+2T_d)}_B$, where $\ket{\psi(3\tau+2T_d)}_A=-i\exp(-i\omega_{PA}(3\tau+2T_d))\ket{e,N'-1}/\sqrt{2}$ and $\ket{\psi(3\tau+2T_d)}_B=-\exp(-i\omega_{PA}(3\tau+2T_d))\ket{g,N'}/\sqrt{2}$. Finally, the last $\pi/2$-pulse causes each of the arms to further split in to $\ket{g,N'}$ and $\ket{e,N'-1}$, so that the state of the system at $t=4\tau+2T_d$ is given by $\ket{\psi(4\tau+2T_d)}=\ket{\psi(4\tau+2T_d)}_A+\ket{\psi(4\tau+2T_d)}_B$, where
\begin{align}
\ket{\psi(4\tau+2T_d)}_A&=\frac{-i}{2}\exp(-i\omega_{PA}(4\tau+2T_d)) \nonumber\\
&\times(-i\ket{g,N'}+\ket{e,N'-1}),\nonumber\\
\ket{\psi(4\tau+2T_d)}_B&=\frac{-1}{2}\exp(-i\omega_{PA}(4\tau+2T_d)) \nonumber\\
&\times(\ket{g,N'}-i\ket{e,N'-1}).
\end{align}
The two arms, thus, yield identical proportions of $\ket{g,N'}$ and $\ket{e,N'-1}$. The probability of finding the atom in the ground state, which is the signal for the CRAIN, is therefore, $S_{CRAIN}=1$. However, if the entire system is rotating at the rate $\Omega_G$ about an axis perpendicular to the area carved by the interferometer, a time delay, $\Delta T$ is introduced between the two paths. To consider its effect on the signal of the CRAIN, we note that the state of the system at $t=3\tau+2T_d$ is such that
\begin{align}
\ket{\psi(3\tau+2T_d)}_A&=\frac{-i}{\sqrt{2}}\exp(-i\omega_{PA}(3\tau+2T_d))\nonumber\\
&\times\exp(i(\omega_{C,g}+\omega_{C,e})\Delta T/4)\ket{e,N'-1},\nonumber\\
\ket{\psi(3\tau+2T_d)}_B&=\frac{-1}{\sqrt{2}}\exp(-i\omega_{PA}(3\tau+2T_d))\nonumber\\
&\times\exp(-i(\omega_{C,e}+\omega_{C,g})\Delta T/4)\ket{g,N'}.
\end{align}
Finally, the state of the system due to rotation at the end of the $\pi/2$-dark-$\pi$-dark-$\pi/2$ sequence is such that
\begin{align}
\ket{\psi(4\tau+2T_d)}_A&=\frac{-i}{2}\exp(-i\omega_{PA}(4\tau+2T_d))\nonumber\\
&\times\exp(i\omega_{C,avg}\Delta T/2)(-i\ket{g,N'}\nonumber\\
&+\ket{e,N'-1}),\nonumber\\
\ket{\psi(4\tau+2T_d)}_B&=\frac{-1}{2}\exp(-i\omega_{PA}(4\tau+2T_d))\nonumber\\
&\times\exp(-i\omega_{C,avg}\Delta T/2)(\ket{g,N'}\nonumber\\
&-i\ket{e,N'-1}),
\end{align}
where $\omega_{C,avg}=(\omega_{C,g}+\omega_{C,e})/2$. The probability of finding the atom in the ground state, which is the signal for the CRAIN, is therefore given by $S_{CRAIN}=\cos^2(\Delta\phi/2)$, where $\Delta\phi=\omega_{C,avg}\Delta T$. From the special relativistic addition of velocities along the two trajectories, the time delay is found to be $\Delta T=2\theta\Omega_G/c^2$, where $\theta$ is the area enclosed by the CRAIN~\cite{Malykin}. In a real experiment, one makes use of a laser, which is a coherent state, and not a Fock state. However, when the mean photon number in the laser is very large, the excitation is akin to what we described here. In effect, the laser in this limit can be viewed effectively as a Fock state with a photon number equaling the mean photon number in the laser. This is the semiclassical approximation, where the quantum state of the field is assumed to remain unchanged (and thus factorized) independent of the state of the atom.

Next, we consider an ensemble of $N$ such two level atoms that are independent and non-interacting. Furthermore, we consider the product state of this ensemble and a Fock state of $N'$ photon as described above. Initially, all the atoms are in the state $\ket{g}$, so that the state of the ensemble-photon system is $\ket{E_0}\ket{N'}\equiv\ket{E_0,N'}$, where $\ket{E_0}=\ket{g_1,g_2,\ldots,g_N}$. Now, let us imagine a scenario (which is impossible in practice) that the state $\ket{E_0,N'}$ is directly coupled to the state $\ket{E_N,N'-N}$ via the exchange of $N$ photons between the states, where $\ket{E_N}=\ket{e_1,e_2,\ldots,e_N}$ as illustrated in~\ref{Fig_2}(b). Such a process can be used to realize an atomic interferometer in a manner analogous to the CRAIN, as illustrated in~\ref{Fig_2}(c). The area enclosed in this case would be the same as that for a CRAIN. However, the average Compton frequency will now be $N\omega_{C,avg}$ (and the de Broglie wavelength will be $\lambda_{dB,SingleAtom}/N$), so that the signal given by the population of state $\ket{E_0}$ measured at the end, will be $S_{ensemble}=\cos^2(N\Delta\phi/2)$, where $\Delta\phi$ is the phase shift experienced by a CRAIN for the same amount of rotation.

\begin{figure}[h]
\includegraphics[scale=0.35]{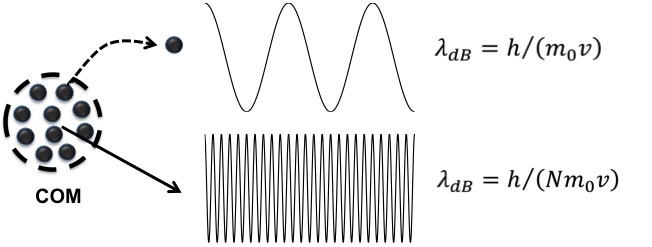}
\caption{\label{Fig_3} $\lambda_{dB}$ of an Rb-87 atom moving at a constant velocity of $1$ m/s is $4.56$ nm). In the rest frame of the atom, its characteristic Compton frequency is $1.96(10^{25}$Hz). A cluster of $10^{6}$ such atoms will exhibit the characteristics of a single entity of mass that is a million times that of a single Rb-87 atom. Therefore, $\lambda_{dB}$ will be $4.56(10^{-15}$ m) and Compton frequency is $1.96(10^{31}$ Hz).}
\end{figure}

However, since the electric dipole moment for a superposition of $\ket{E_0}$ and $\ket{E_N}$ vanishes, there is no way to realize the type of excitation envisioned above. Instead, when excited by a Fock state of $N'(>N)$ photons, this ensemble unfolds into a superposition of $(N+1)$ symmetric collective states given by $\ket{E_n}\ket{N'-n}$, where $\ket{N'-n}$ is a state of the field with $(N'-n)$ photons and $\ket{E_n}=J(N,n)^{-1/2}\sum_{k=1}^{J(N,n)}P_k\ket{g^{\otimes(N-n)}e^{\otimes n}}$, $J(N,n)={N \choose n}$, $P_k$ is the permutation operator, and $n=0,1,2,\ldots,N$ \cite{HumeWineland}. The state $\ket{E_n}$ has a momentum of $n\hbar{k}$ in the direction of the beam, since it has absorbed $n$ photons. Thus, it will exhibit collective behavior such that its center of mass (COM) recoils with a velocity equal to $n\hbar{k}/Nm$. As such, an ensemble in such a state can be viewed as a single entity with a mass of $Nm$ and a Compton frequency $\omega_C$ that is $N$ times that of a single constituent atom, despite no interaction between the atoms. Conversely, the ensemble can also be viewed as having a de Broglie wavelength $\lambda_{dB}=h/Nmv$ that is $N$ times smaller than that of a single atom, where $v$ is the magnitude of the total velocity (e.g., a constant velocity in the \textbf{x}-direction that is much larger than the recoil velocity). This is illustrated schematically in Fig~\ref{Fig_3}.

\begin{figure}[h]
\centering
\includegraphics[scale=0.25]{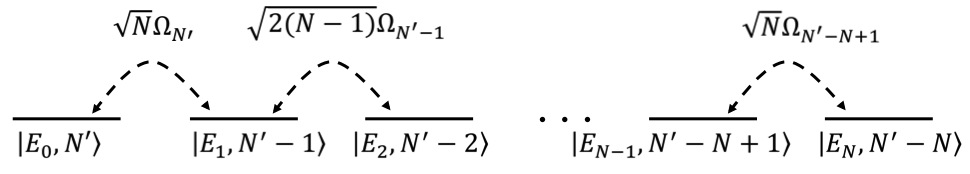}
\caption{\label{Fig_4} Coupling between an $N$-atom ensemble symmetric collective states and $N'$ photons.}
\end{figure}

Some of these states and their relevant couplings are illustrated in Fig.~\ref{Fig_4}. For example, state $\ket{E_0,N'}$ is coupled to the state $\ket{E_1,N'-1}$ at the rate of $\sqrt{N}\Omega_{N'}$, where $\Omega_{N'}=\sqrt{N'}\Omega_0$, with $\Omega_0$ being the single-photon Rabi frequency (for exciting a single atom) and the $\sqrt{N}$ factor results from the collective enhancement of coupling. If the excitation is carried out by a laser field where the mean photon number is much larger than $N$, then we can make a semiclassical approximation that $\Omega_{N'}\cong\Omega_{N'-1}\cong\ldots\cong\Omega_{N'-N}\equiv\Omega$. Furthermore, the quantum state of the laser remains unchanged, (and thus factorized) independent of the state of the ensemble. The Compton frequency of the state $\ket{E_n}$ is given by $\omega_{C,En}=m_{En}c^2/\hbar$, where $m_{En}=m_{E0}+n\hbar\omega/c^2$ is the rest mass of the ensemble in state $\ket{E_n}$, and $m_{E0}=Nm$ is the rest mass of the ensemble in state $\ket{E_0}$. Thus, the dressed frequency of the ensemble-photon system $\omega_{PE}$, which is a constant for all possible states of the system can be written as $\omega_{PE}(N,N')=m_{En}c^2/\hbar+(N'-n)\omega=m_{E0}c^2/\hbar+N'\omega$.

\begin{figure}[h]
\centering
\includegraphics[scale=0.40]{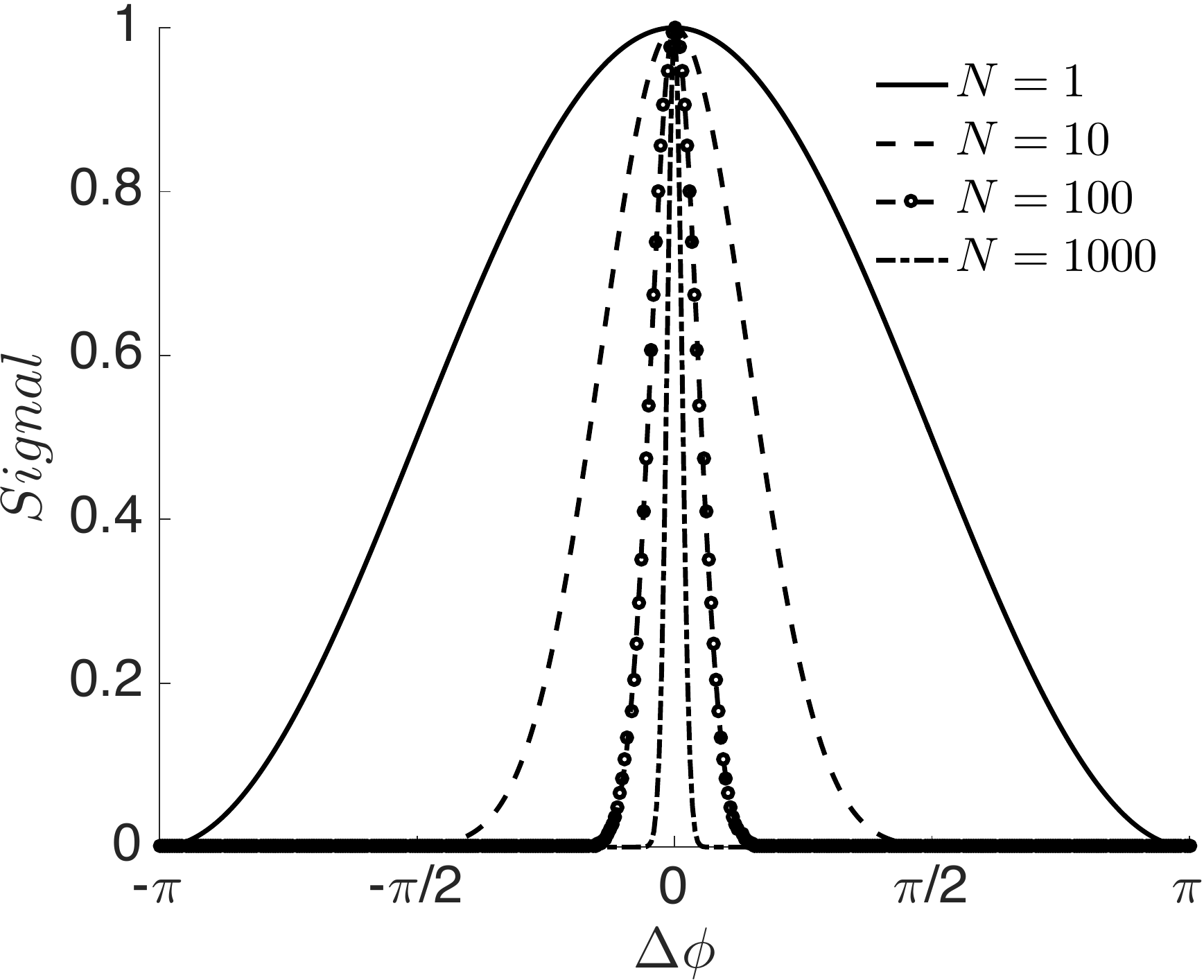}
\caption{\label{Fig_5} Measurement of the COSAIN signal (amplitude of $\ket{E_0}$)
shows a narrowing of the fringe width such that the ratio $\varrho(1)/\varrho(N)$ increases with $\sqrt{N}$.}
\end{figure}

\begin{figure}[h]
\centering
\includegraphics[scale=0.5]{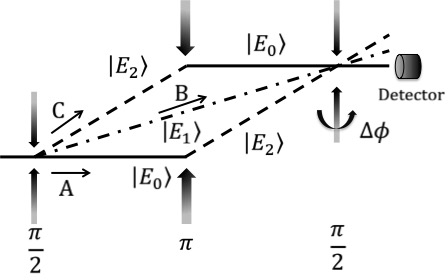}
\caption{\label{Fig_6}Illustration of a two atom COSAIN depicting the state trajectories.}
\end{figure}

In the absence of an effective detuning, the COSAIN is based on the coherent splitting and recombining of all of these symmetric collective states. The signal of the COSAIN is, thus, the product of the signals of the constituent CRAIN's that work simultaneously, resulting in the narrowing of the signal fringes. The fringe linewidth, defined as the full width at half maximum (FWHM) of the signal fringe is given by $\varrho(N) = 2\cos^{-1}(2^{-1/2N})$. It is evident from Fig.~\ref{Fig_5} that the $\varrho(N)$ decreases with increasing $N$. To illustrate the mechanism behind the COSAIN more transparently, we now consider the simplest ensemble: an assembly of two atoms of the kind described above and $N'$ photons. At $t=0$, the ensemble-photons system is assumed to be in the state $\ket{E_0,N'}$. The atom-field interaction couples it to the state $\ket{E_1,N'-1}$, which in turn is coupled to the state $\ket{E_2,N'-2}$. Following the notations of the $\pi/2$-dark-$\pi$-dark-$\pi/2$ sequence established for the CRAIN, the state of the ensemble after the first $\pi/2$ pulse is $\ket{\Psi(\tau)}=\exp(-i\omega_{PE}\tau)(\ket{E_0,N'}_A-i\sqrt{2}\ket{E_1,N'-1}_B-\ket{E_2,N'-2}_C)/2$, where $\omega_{PE}\equiv\omega_{PE}(2,N')$ and the subscripts $A$, $B$, and $C$ denote the lower, middle and upper trajectories of the interferometer, respectively, as shown in Fig.~\ref{Fig_6}. This is followed by a dark zone of duration $T_d$, at the end of which the state of the ensemble is $\ket{\Psi(\tau+T_d)}=\exp(-i\omega_{PE}T_d)\ket{\Psi(\tau)}$. The component $\ket{E_1,N'-1}_B$ is displaced by $\hbar kT_d/2m$ along the \textbf{z}-axis since it has absorbed the recoil from one photon $(\hbar k)$, and it has a mass of $2(m_{C,g}+m_{C,e})\approx 2m$. Similarly, $\ket{E_2,N'-2}$ is displaced by $\hbar kT_d/m$ along the \textbf{z}-axis since it has absorbed recoils from two photons $(2\hbar k)$, and it has a mass of $2m_{C,e}\approx 2m$. At $t=\tau+T_d$, the system interacts with the $\pi$-pulse (of duration $2\tau$) which causes the transition $\ket{E_0,N'}\leftrightarrow\ket{E_2,N'-2}$. The state $\ket{E_1,N'-1}$, however only picks up a phase due to the $\pi$ interaction, and its trajectory remains unchanged. Explicitly, the state of the system at the end of the $\pi$-pulse is $\ket{\Psi(3\tau+T_d)}=\ket{\Psi(3\tau+T_d)}_A+\ket{\Psi(3\tau+T_d)}_B+\ket{\Psi(3\tau+T_d)}_C$, where
\begin{align}
\ket{\Psi(3\tau+T_d)}_A &= -\frac{1}{2}\exp(-i\omega_{PE}(3\tau+T_d))\ket{E_2,N'-2},\nonumber\\
\ket{\Psi(3\tau+T_d)}_B &= -\frac{1}{\sqrt{2}}\exp(-i\omega_{PE}(3\tau+T_d))\nonumber\\
&\times\ket{E_1,N'-1},\nonumber\\
\ket{\Psi(3\tau+T_d)}_C &= \frac{1}{2}\exp(-i\omega_{PE}(3\tau+T_d))\ket{E_0,N'}.
\end{align}

At the end of this pulse, the system passes through a second dark zone of duration $T_d$, which causes the state of the system to become $\ket{\Psi(3\tau+2T_d)}=\ket{\Psi(3\tau+2T_d)}_A+\ket{\Psi(3\tau+2T_d)}_B+\ket{\Psi(3\tau+2T_d)}_C$, where $\ket{\Psi(3\tau+2T_d)}_A=\exp(-i\omega_{PE}T_d)\ket{\Psi(3\tau+T_d)}_A$, $\ket{\Psi(3\tau+2T_d)}_B=\exp(-i\omega_{PE}T_d)\ket{\Psi(3\tau+T_d)}_B$, and $\ket{\Psi(3\tau+2T_d)}_C=\exp(-i\omega_{PE}T_d)\ket{\Psi(3\tau+T_d)}_C$. By the end of this dark zone, the three trajectories converge and a last $\pi/2$-pulse is applied which causes each of the trajectories to further split as follows:
\begin{align}
\ket{\Psi}_A &= \frac{-1}{4}\exp(-i\omega_{PE}(4\tau+2T_d))(-\ket{E_0,N'}\nonumber\\
&-i\sqrt{2}\ket{E_1,N'-1}+\ket{E_2,N'-2}),\nonumber\\
\ket{\Psi}_B &= \frac{1}{2}\exp(-i\omega_{PE}(4\tau+2T_d))(\ket{E_0,N'}\nonumber\\
&+\ket{E_2,N'-2}),\nonumber\\
\ket{\Psi}_C &= \frac{1}{4}\exp(-i\omega_{PE}(4\tau+2T_d))(\ket{E_0,N'}\nonumber\\
&-i\sqrt{2}\ket{E_1,N'-1}-\ket{E_2,N'-2}).
\end{align}

The signal of the COSAIN is the probability of finding the ensemble in any of the collective states. We choose to measure the probability of $\ket{E_0,N'}$. The probability of finding the ensemble in state $\ket{E_0,N'}$, is therefore, $S_{COSAIN}=1$. However, as explained above for the case of the CRAIN, if the entire system is rotating at the rate $\Omega_G$ about an axis perpendicular to the area carved by the interferometer, a time delay is introduced between the paths. This time delay depends only on the area enclosed and the rate of rotation, as noted earlier. Let us assume that the delay between the paths $C$ and $A$, which forms the $A-C$ loop, is $\Delta T$. Therefore, the delay between paths $B$ and $A$ which form the $A-B$ loop, will be $\Delta T/2$. Similarly, the delay between paths $C$ and $B$, which form the $B-C$ loop, will also be $\Delta T/2$. Since only the relative delay between two paths matter, we assume, for simplicity, that there is no delay on path $B$. Thus, just before the final $\pi/2$ pulse, we can write the quantum states of these paths under rotation as $\ket{\Psi}_{BR}=\ket{\Psi(3\tau+2T_d)}_B$, $\ket{\Psi}_{AR}=\exp(i(\omega_{C,E0}+\omega_{C,E2})\Delta T/4)\ket{\Psi(3\tau+2T_d)}_A$, and $\ket{\Psi}_{CR}=\exp(-i(\omega_{C,E2}+\omega_{C,E0})\Delta T/4)\ket{\Psi(3\tau+2T_d)}_C$. The last $\pi/2$-pulse causes each of these components to further split so that the state of the system at the end of the $\pi/2$-dark-$\pi$-dark-$\pi/2$ sequence is
\begin{align}
\ket{\Psi}_{AR} &= \frac{-1}{4}\exp(-i\omega_{PE}(4\tau+2T_d))\nonumber\\
&\times\exp(i(\omega_{C,E0}+\omega_{C,E2})\Delta T/4)\nonumber\\
& \times (-\ket{E_0,N'}-i\sqrt{2}\ket{E_1,N'-1}+\ket{E_2,N'-2}),\nonumber\\
\ket{\Psi}_{BR} &= \frac{1}{2}\exp(-i\omega_{PE}(4\tau+2T_d))(\ket{E_0,N'}\nonumber\\
&+\ket{E_2,N'-2}),\nonumber\\
\ket{\Psi}_{CR} &= \frac{1}{4}\exp(-i\omega_{PE}(4\tau+2T_d))\nonumber\\
&\times\exp(-i(\omega_{C,E2}+\omega_{C,E0})\Delta T/4)\nonumber\\
& \times (\ket{E_0,N'}-i\sqrt{2}\ket{E_1,N'-1}-\ket{E_2,N'-2}).
\end{align}

\begin{figure}[h]
\centering
\includegraphics[scale=0.35]{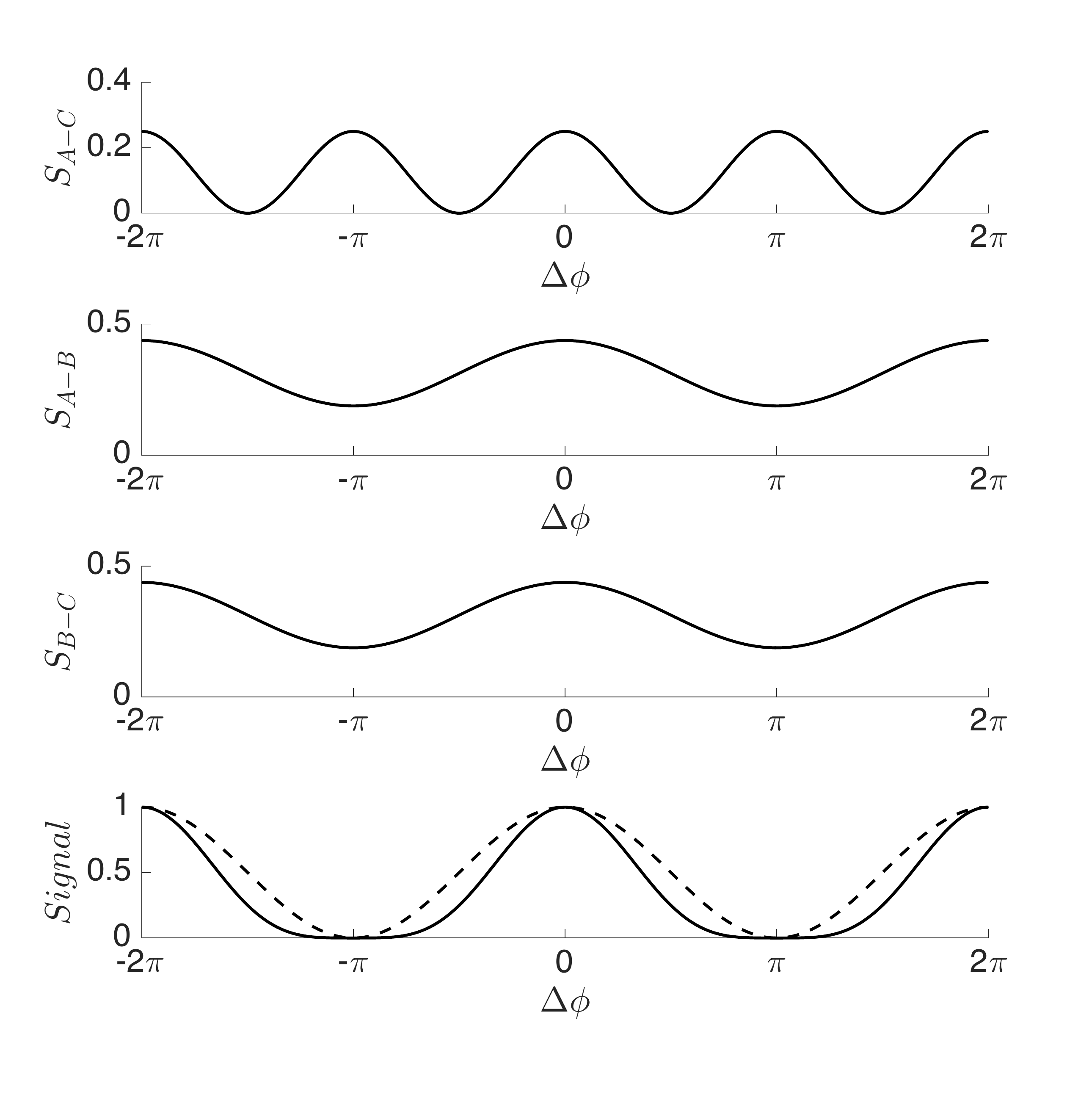}
\caption{\label{Fig_7}Signals derived from the interferometers formed by trajectories $A-C$, $A-B$, and $B-C$. The bottom panel shows the signal of CRAIN (broken line) to the signal of a 2-atom COSAIN (solid line).}
\end{figure}

The signal of the COSAIN can, thus, be viewed as the aggregation of interference patterns due to three independent CRAIN's working simultaneously., i.e. those formed by paths $A-B$, $B-C$ and $A-C$. To illustrate this, we denote the component of $\ket{E_0,N'}$ in paths $A$, $B$ and $C$ as $\chi_A$, $\chi_B$ and $\chi_C$, respectively. The interferometers formed by $A-B$ and $B-C$ are identical. The measurement of the amplitude of $\ket{E_0,N'}$ from each of these interferometers is given by $S_{A-B}=S_{B-C}=|\chi_A+\chi_B|^2=|\chi_B+\chi_C|^2=3/16+\cos^2(\omega_{C,avg}\Delta T/2)/4$. This corresponds to a CRAIN that is operating with an atom of average Compton frequency $\omega_{C,avg}$. The interferometer formed by $A-C$ yields the signal value $S_{A-C}=|\chi_A+\chi_C|^2= \cos^2(\omega_{C,avg}\Delta T)/4$, behaving analogously to a CRAIN formed by an atom of average Compton frequency $2\omega_{C,avg}$. The total COSAIN signal arises due to the interference of the component of $\ket{E_0,N'}$ from the three paths, $S_{COSAIN}=|\chi_A+\chi_B+\chi_C|^2=\cos^4(\omega_{C,avg}\Delta T/2)$, as shown in Fig~\ref{Fig_7}. This is reconciled by the fact that $|\chi_A+\chi_B+\chi_C|^2=|\chi_A+\chi_B|^2+|\chi_B+\chi_C|^2+|\chi_A+\chi_C|^2-(|\chi_A|^2+|\chi_B|^2+|\chi_C|^2)$. The collective atomic recoil laser (CARL) mechanism is similar to this concept presented here, in the sense that no interaction between atoms are needed \cite{CARL,CARLHemmer}. On the other hand, the Dicke Phase transition pertains to the BEC regime, and is not closely related to what is being presented here\cite{DickePhase}.

\section{\label{sec:inhomogeneities}Parameter Inhomogeneities Affecting Signal}

In this section, we present a detailed description of the effect of inhomogeneity in Rabi frequency and Doppler shift on the signal of a COSAIN. These inhomogeneities put significant constraints on the ensemble size, temperature of the trapped atoms, and the intensity profile and size of the laser beams. The manifestations of these effects can be analyzed by considering an ensemble of $N$ identical non-interacting and independent atoms of the type described in Sec.~\ref{sec:narrowing}. A laser beam propagating along the \textbf{z}-axis will impart a momentum $\hbar{k}$ to an atom upon absorption of recoil from a single photon, driving it to a superposition of the states $\ket{g_i,0}$ and $\ket{e_i,\hbar k}$, with the amplitude of each state depending on the intensity of the laser beam and the time of interaction. The field amplitude of the laser beams are assumed to be of Gaussian profile in \textbf{x} and \textbf{y} directions, and constant in the \textbf{z} direction. At $t=0$, the position of the $i$-th atom is given by $\mathbf{\vec{r}}=x_i\mathbf{\hat{x}}+y_i\mathbf{\hat{y}}+z_i\mathbf{\hat{z}}$. Due to the thermal motion of the atoms, each atom experiences a different Doppler shift  and therefore, a different effective laser frequency, $\omega_{0i}$. The net consequence of this is that the $i$-th atom picks up a detuning of $\delta_i = k\mathbf{v_{iz}}$, where $\mathbf{v_{iz}}$ denotes the atom's velocity in the \textbf{z}-direction. Furthermore, each atom sees a different electric field,  $\mathbf{E_i}=\mathbf{\hat{x}}E_0\exp(-(x_i^2+y_i^2)/2/\sigma_L^2)\cos(\omega_{0i}t-kz_i)$, due to the finite extent of the ensemble. Here $\sigma_L$ represents the width of the laser beam in the transverse directions. Therefore, the Rabi frequency experienced by the $i$-th atom is given by $\Omega_i=\Omega_0\exp[-(x^2+y^2)/2\sigma_L^2]$, where $\Omega_0\equiv\bra{g_i}(\mathbf{x}\cdot\mathbf{\rho_i})\ket{e_i}E_0/\hbar$ $=\bra{e_i}(\mathbf{x}\cdot\mathbf{\rho_i})\ket{g_i}E_0/\hbar$ and $\mathbf{\rho}_i$ is the position of the electron with respect to the nucleus.

In the electric dipole approximation, the Hamiltonian for the $i$-th atom can be written as $H_i=|{\mathbf{p}_{i}}|^{2}\slash2m+ H_{0i}+q\mathbf{\rho_i.E_i}$, where $H_{0i}$ is the internal energy of the atom, $q$ is the electronic charge, $m$ is the mass of the atom, and $\mathbf{p_i}$ is the momentum of the $i$-th atom. The COM motion kinetic energy term be expressed as $|{\mathbf{p}_{i}}|^{2}\slash2m=|{\mathbf{p}_{iz}}|^{2}\slash2m+|{\mathbf{p}_{i\perp}}|^{2}\slash2m$, where $\mathbf{p}_{iz}$ is the momentum in the \textbf{z} direction, and $\mathbf{p}_{i\perp}$ is the momentum in a direction perpendicular to \textbf{z}. Consider first the effect of the second term: $|{\mathbf{p}_{i\perp}}|^{2}\slash2m$. In a typical experimental scenario, this accounts for the motion of the atom, typically at a large velocity, in the \textbf{x}-direction (see Fig.~\ref{Fig_1}), acquired, for example by the initial push imparted to the trapped atoms before they enter the first interaction zone. Thus, any variation in this due to a velocity spread within the ensemble can be ignored, and this term can be treated as an overall constant energy which can be subtracted from the Hamiltonian. Consider next the first term: $|{\mathbf{p}_{iz}}|^{2}\slash2m$. This term shows that the state $\ket{g,\mathbf{p_{iz}}}$ coupled to $\ket{e,\mathbf{p_{iz}}+\hbar k}$ by the laser differ in energy by $(\hbar kv_{iz}+\hbar^2k^2/2m)$, where the first term is the Doppler shift and the second term is the recoil energy which is a constant for all atoms, and can be subtracted from the Hamiltonian. Thus, after subtraction of constant terms, the net effect of the kinetic energy term is to account for the Doppler shift. Finally, as we have shown in detail in ref.~\citenum{CollectiveDescription}, a fully quantum mechanical description of the COM motion (e.g., by keeping track explicitly of the momentum of the atoms in the $\ket{g}$ and $\ket{e}$ states) is not essential in describing the collective states in the limit where the Rabi frequency of the $i$-th atom, $\Omega_i$, is large compared to the Doppler shift due to the COM motion. This regime is valid for the COSAIN, and, therefore, a semiclassical description of the COM motion of each atom suffices for the case at hand. Upon making the rotating-wave approximation, $H_i$ can then be expressed in the bases of $\ket{g_i}$ and $\ket{e_i}$ as $H_i/\hbar =\omega_g\ket{g_i}\bra{g_i}+\omega_e\ket{e_i}\bra{e_i}+ \Omega_i(\exp(i(\omega_{0i}t-kz_i))\ket{g_i}\bra{e_i}+ h.c.)/2$, where $\omega_e$ includes the Doppler shift. Performing the rotating-wave transformation and removing any phase factors causes the transformation $H_i\rightarrow H_i'$, such that  $H_i'/\hbar = -\delta_i\ket{e_i'}\bra{e_i'}+\Omega_i(\ket{g_i'}\bra{e_i'}+ h.c.)/2$. The new basis vectors, $\ket{g_i'}$ and  $\ket{e_i'}$, are related to the original basis vectors as $\exp(-i\omega_gt)\ket{g_i}$ and $\exp(-i((\omega_e+\delta_i)t-kz_i))\ket{e_i}$, respectively. Assuming that the $i$-th atom is initially in the state $c_{gi}(0)\ket{g_i'}+c_{ei}(0)\ket{e_i'}$, its quantum state can be written as 
\begin{align}
\ket{\psi_i'} =&  e^{i\delta_i t/2}((c_{gi}(0)\cos\left(\frac{\Omega_i't}{2}\right)\nonumber\\
&-i\frac{c_{gi}(0)\delta_i+c_{ei}(0)\Omega_i}{\Omega_i'}\sin\left(\frac{\Omega_i't}{2}\right))\ket{g_i'}\nonumber \\
& +(-i\frac{c_{gi}(0)\Omega_i-c_{ei}(0)\delta_i}{\Omega_i'}\sin\left(\frac{\Omega_i't}{2}\right)\nonumber\\
&+c_{ei}(0)\cos\left(\frac{\Omega_i't}{2}\right))\ket{e_i'}),
\label{eq_SA_evolution_general}
\end{align}
where $\Omega_i'  = \sqrt{\Omega_i^2+\delta_i^2}$ is the effective coupling frequency of this atom. The relative separation of the atoms along the direction of propagation of the laser beam has no effect on the fidelity of the collective states that can be attained by the ensemble~\cite{CollectiveDescription}. For the purpose of the present discussion, we stay in the bases of $\ket{g_i'}$ and $\ket{e_i'}$.

At $t=0$, the first pulse of duration $\tau$ is applied to the atoms so that $\Omega_0\tau=\pi/2$. The state of the $i$-th atom following this interaction can be written as $\ket{\psi_i'(\tau)}= c_{gi}(\tau)\ket{g_i'}_A+c_{ei}(\tau)\ket{e_i'}_B$, where $c_{gi}(\tau)= \exp(i\delta_i \tau/2)((\cos\left(\Omega_i'\tau/2\right)-i\delta_i\sin\left(\Omega_i'\tau/2\right)/\Omega_i')$ and $c_{ei}(\tau)=\exp(i\delta_i \tau/2)(-i\Omega_i\sin\left(\Omega_i'\tau/2\right)/\Omega_i')$. The subscripts $A$ and $B$ denote the lower and upper arm of the interferometer trajectory. The ensuing dark zone lasts for a duration $T_d$ wherein the atoms are left to drift freely so that at $t=\tau+T_d$, the COM of state $\ket{e_i'}$ is separated from that of state $\ket{g_i}$ by $d=\hbar kT_d/m$. During this dark zone where no atom-light interaction is taking place, the portion of the atom in state $\ket{e_i'}$ picks up a phase due to detuning, making the state of the atom at the end of this pulse $\ket{\psi_i'(\tau+T_d)} = c_{gi}(\tau+T_d)\ket{g_i'}_A+c_{ei}(\tau+T_d)\ket{e_i'}_B$, where $c_{gi}(\tau+T_d)=c_{gi}(\tau)$ and $c_{ei}(\tau+T_d)=\exp(i\delta_iT_d)c_{ei}(\tau)$. At this point a second pulse of duration $2\tau$ ($\pi$-pulse) is applied to atoms, and each trajectory undergoes further splitting, as shown in Fig.~\ref{Fig_1}. The $\pi$-pulse can, in principle, be perfect only for one group of atoms, such as those with $\delta=0$. For all other atoms, the pulse duration will differ slightly from $\pi$. As a result, for example, the $\ket{e_i'}$ state will not fully evolve into the $\ket{g_i'}$ state, and a residual amount will stay in the $\ket{e_i'}$ state. In the regime where $\Omega_i\gg\delta_i$ for all $i$, the effect of these residual components can be safely ignored. Under this approximation, the state of the atom is given by $\ket{\psi_i'(3\tau+T_d)} = c_{ei}(3\tau+T_d)\ket{e_i'}_A+c_{gi}(3\tau+T_d)\ket{g_i'}_B$, where $c_{ei}(3\tau+T_d)=\exp(i\delta_i\tau)c_{gi}(\tau+T_d)(-i\Omega_i\sin(\Omega_i'\tau)/\Omega_i')$ and $c_{gi}(3\tau+T_d)=\exp(i\delta_i\tau)c_{ei}(\tau+T_d)(-i\Omega_i\sin(\Omega_i'\tau)/\Omega_i')$. Following the $\pi$-pulse, the atoms are further set adrift in another dark zone of duration $T_d$, where the component of the atom following trajectory $A$ picks up a phase due to detuning. The net effect of this is that $\ket{\psi_i'(3\tau+2T_d)} = c_{ei}(3\tau+2T_d)\ket{e_i'}_A+c_{gi}(3\tau+2T_d)\ket{g_i'}_B$, where $c_{ei}(3\tau+2T_d)=\exp(i\delta_iT_d)c_{ei}(3\tau+T_d)$ and $c_{gi}(3\tau+2T_d)=c_{ei}(3\tau+T_d)$. By the end of this dark zone, the two trajectories converge and a third pulse of duration $\tau$ is applied to the atoms. Therefore, the state of the atom at $t=4\tau+2T_d$ is $\ket{\psi_i'(4\tau+2T_d)}=(c_{gi}(4\tau+2T_d)_A\ket{g_i'}+c_{ei}(4\tau+2T_d)_A\ket{e_i'})+(c_{gi}(4\tau+2T_d)_B\ket{g_i'}+c_{ei}(4\tau+2T_d)_B\ket{e_i'})$, where $c_{gi}(4\tau+2T_d)_A=\exp(i\delta_i\tau/2)c_{ei}(3\tau+2T_d)(-i\Omega_i\sin(\Omega_i'\tau/2)/\Omega_i')$, $c_{ei}(4\tau+2T_d)_A=\exp(i\delta_i\tau/2)c_{ei}(3\tau+2T_d)(\cos(\Omega_i'\tau/2)+i\delta_i\sin(\Omega_i'\tau/2)/\Omega_i')$, $c_{gi}(4\tau+2T_d)_B=\exp(i\delta_i\tau/2)c_{gi}(3\tau+2T_d)(\cos(\Omega_i'\tau/2)-i\delta_i\sin(\Omega_i'\tau/2)/\Omega_i')$, and $c_{ei}(4\tau+2T_d)_B=\exp(i\delta_i\tau/2)c_{gi}(3\tau+2T_d)(-i\Omega_i\sin(\Omega_i'\tau/2)/\Omega_i')$.

The signal of the CRAIN formed by the $i$-th atom is the measurement of the amplitude of state $\ket{g_i'}$ at the end of the $\pi/2$-dark-$\pi$-dark-$\pi/2$ sequence due to the interference of the components from the two paths. Since the two arms yield identical proportions of both $\ket{g_i'}$ and $\ket{e_i'}$, i.e. $c_{gi}(4\tau+2T_d)_A=c_{gi}(4\tau+2T_d)_B$ and $c_{ei}(4\tau+2T_d)_A=-c_{ei}(4\tau+2T_d)_B$, the signal of the CRAIN formed is $S_{CRAIN,i}=\alpha_i$, where $\alpha_i=|2c_{gi}(4\tau+2T_d)_A|^2\leq 1$. Since the signal of a COSAIN is the product of the signals of the individual CRAIN's formed by the constituent atoms in the ensemble~\cite{CollectiveDescription}, the signal of the resulting COSAIN is, consequently, $S_{COSAIN}=\prod_i^NS_{CRAIN,i}=\prod_i^N\alpha_i$. However, if a phase difference is introduced between the two paths, the signal of the CRAIN's and thus, the COSAIN will depend on it additionally. Assuming that an external phase, $\Delta\phi$ is introduced to the path $A$ of the interferometer, the quantum state of the atom at $t=4\tau+2T_d$ is given by $\ket{\psi_i'(4\tau+2T_d)}=\exp(i\Delta\phi)(c_{gi}(4\tau+2T_d)_A\ket{g_i'}+c_{ei}(4\tau+2T_d)_A\ket{e_i'})+(c_{gi}(4\tau+2T_d)_B\ket{g_i'}+c_{ei}(4\tau+2T_d)_B\ket{e_i'})$. The amplitude of $\ket{g_i'}$ will, thus, be $S_{CRAIN,i}=|1+\exp(-i\Delta\phi)|^2\alpha_i=4\alpha_i\cos^2(\Delta\phi/2)$. In the case where $\Omega_i\gg\delta_i$, $\alpha_i =1/4$ and the signal shows the well known $\cos^2(\Delta\phi/2)$ dependence. The resulting COSAIN signal is, therefore, $S_{COSAIN}=\prod_i^N4\alpha_i\cos^2(\Delta\phi/2)$. In the ideal situation where each atom sees the same Rabi frequency due to a uniform beam profile and there is no effective detuning experienced by the atoms, $\alpha_i=1/4$ and the signal at the end of the interferometer sequence is given by $S_{COSAIN}=\cos^{2N}(\Delta\phi/2)$. This corresponds to the narrowing of the signal fringe by a factor proportional to $\sqrt{N}$ as compared to the signal in a CRAIN. 

In the more practical situation relevant for experimental conditions, $\Omega_i$ and $\delta_i$ for each atom are determined by the laser beam intensity profile, and atom trap size and temperature, as described above. To illustrate the effect of these parameters, we assume that the atoms are first cooled down using a magneto-optic trap arrangement. The trapped atoms are then held in a cigar-shaped dipole trap to further cool them down via evaporative cooling. The density of atoms in the trap is assumed to follow a Gaussian spatial distribution so that its length is in the longitudinal direction is $\xi_L$, and its width in the transverse direction is $\xi_T$.

\subsection{\label{subsec:DopplerShift}Effect of Velocity Distribution}

\begin{figure}[h]
\includegraphics[scale=0.35]{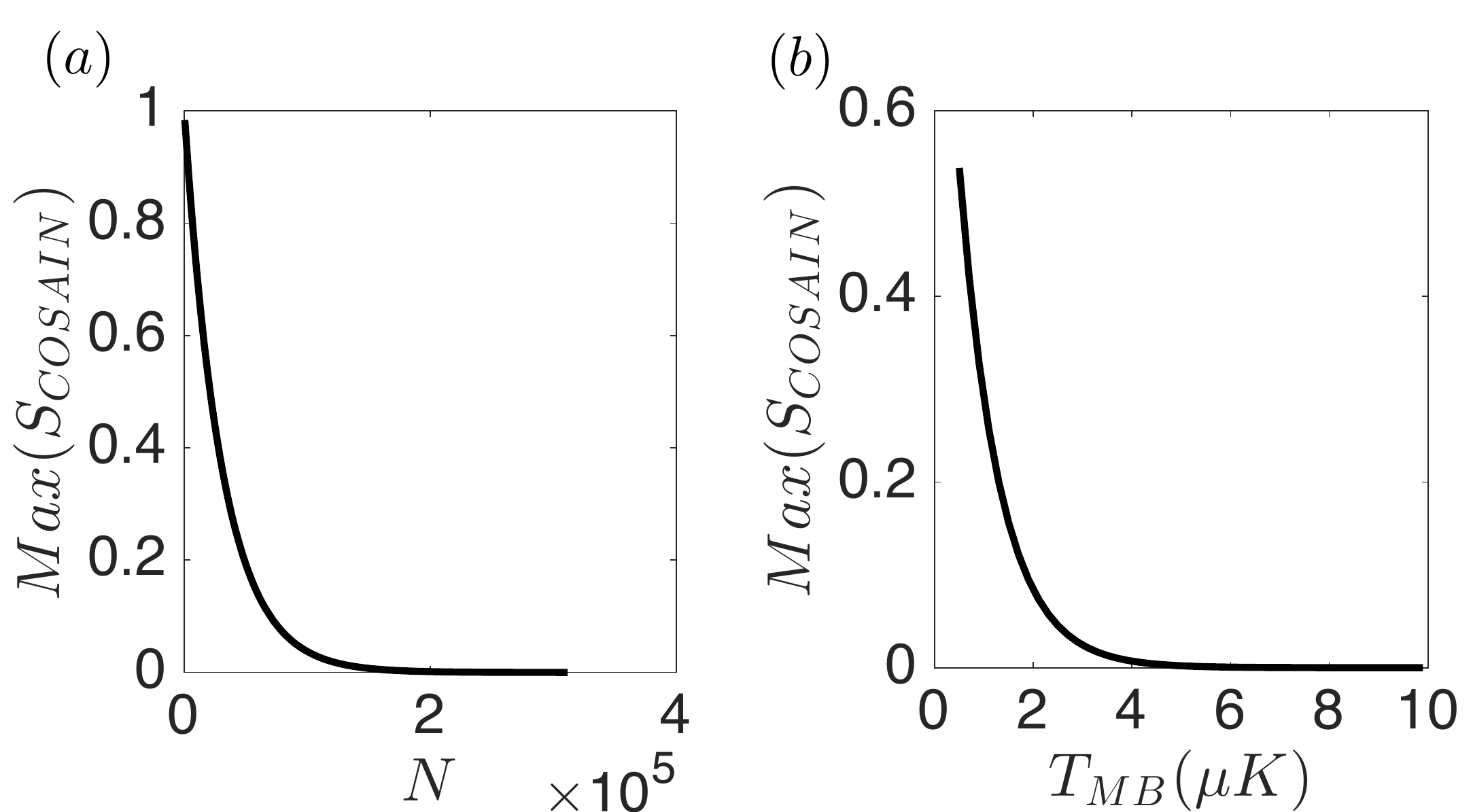}
\caption{\label{Fig_8}(a) Variation of signal peak value with $N$ at $0.5\mu$K average temperature and rectangular intensity profile beams at $\Omega = 1.9\times 10^7$s$^{-1}$. (b) Variation of signal peak value with trap temperature for $N\simeq 1.9\times 10^4$.}
\end{figure}

The Maxwell-Boltzmann velocity distribution of the ensemble is $f_{MB}(v,T_{MB})=\sqrt{m/2\pi k_BT_{MB}}$ $\times\exp(-mv^2/2\pi k_BT_{MB})$, where $T_{MB}$ is the temperature of the trap and $k_B$ is the Boltzmann constant. Since the ensemble undergoes interaction with a pair of counter-propagating laser beams, the Doppler shift observed by the $i$-th atom, $\delta_i=(k_1+k_2)v_i$ cannot be neglected compared to the Raman-Rabi frequency experienced by it. Thus, at non-zero ensemble temperatures, the signal contribution from each atom is significantly lower than the maximum amplitude possible. The signal peak value falls sharply with increasing $N$ as illustrated in Fig~\ref{Fig_8}(a). It is also evident from Fig.~\ref{Fig_8}(b) that the signal of a COSAIN varies significantly as a function of the temperature.

\subsection{\label{subsec:LaserProfile}Effect of Intensity Profile of Laser Beams}

\begin{figure}[h]
\includegraphics[scale=0.35]{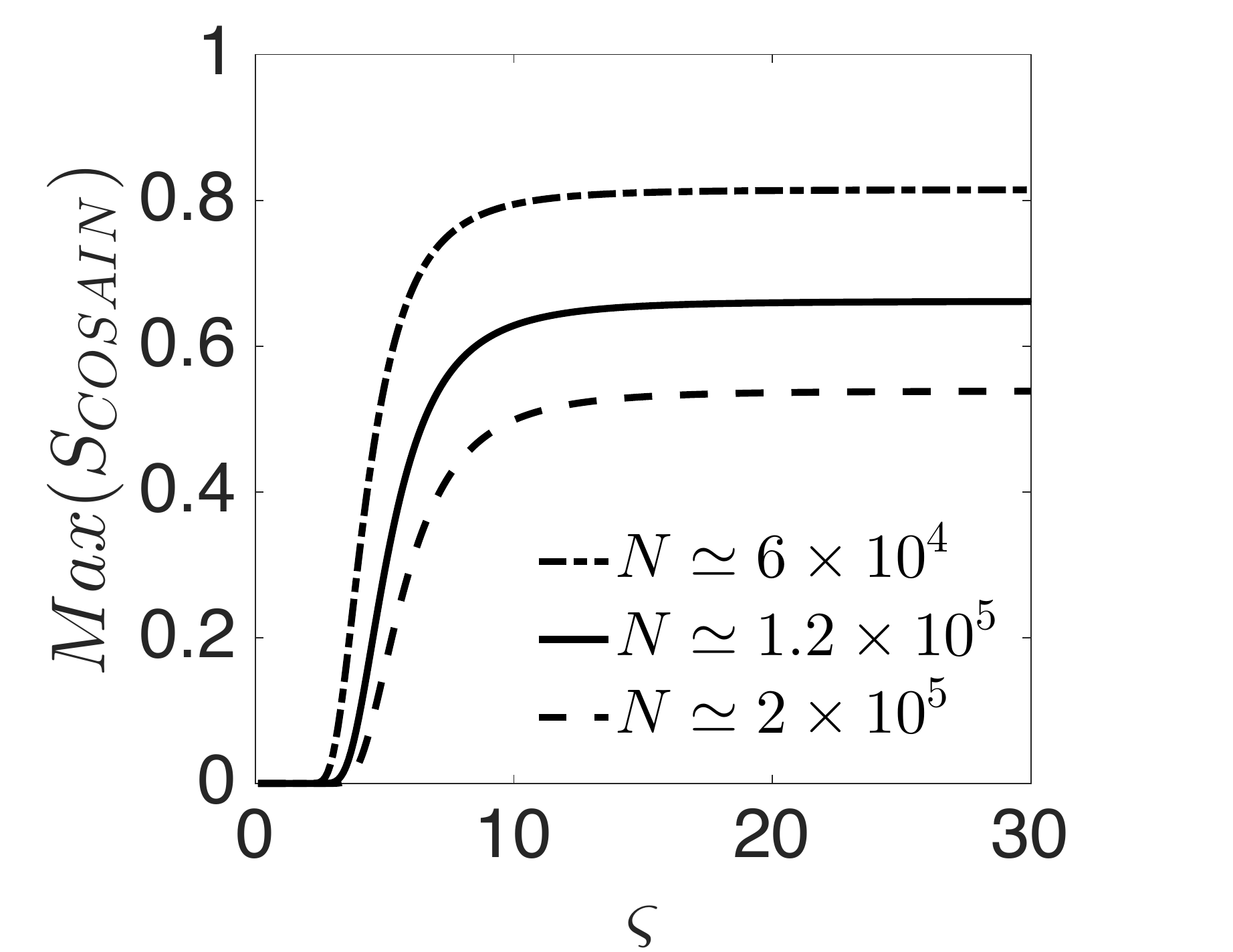}
\caption{\label{Fig_9} Variation of the peak value of the $S_{COSAIN}$ with increasing MOT size to beam waist ratio at $T_{MB}=0.5\mu$K for different values of $N$. }
\end{figure}

Next, we consider the effect of the Gaussian spatial distribution of the Raman beams on the COSAIN. Assuming that the beam waist size is $w$, the Raman Rabi frequency experienced by the $i$-th atom of the ensemble is $\Omega_i=\Omega_0\exp(-2r^2/w^2)$. Here $\Omega_0$ is the peak value of the Raman Rabi frequency and $r$ is the radial distance of the $i$-th atom from the center of the beam. We consider that the average temperature of the trapped atoms is $T_{MB}=0.5\mu$K and the peak value of the beam intensity is $15$ mW/mm$^2$ so that $\Omega_0=1.9\times 10^7$ rads$^{-1}$. Figure~\ref{Fig_9} shows the variation of the peak value of the $S_{COSAIN}$ with increasing value of $\varsigma=w/\xi_T$.

\subsection{\label{subsec:spontaneous_emmision}Effect of spontaneous emission}

In our analysis of the COSAIN, we have employed a model of a three-level atom where the intermediate state ($\ket{a,\hbar k_1}$) is adiabtically eliminated to reduce the system to an equivalent two-level model. However, the actual population of this state is approximately $\Omega^2/\delta^2$, with $\Omega=\Omega_1=\Omega_2$. In the time that it takes for a $2\pi$ pulse ($\pi/2-\pi-\pi/2$ sequence sans the dark zones), we can estimate that the number of spontaneous emissions that occur per atom is $2(\Omega^2/\delta^2)\tau\Gamma\simeq4\pi\Gamma/\delta$. For $\delta=200\Gamma$, this number is about $6.3\times10^{-2}$ and increases by a factor of $N$ for an ensemble of $N$ atoms. Note that there is no enhancement in the rate of spontaneous emission due to superradiant effects, since we are considering a dilute ensemble. Consequently, the signal for both the CRAIN and the COSAIN would deviate from the ideal one. The effect of spontaneous emission on the CRAIN can be taken into account by using the density matrix equation for a three level system. However, in this case, it is not possible to ascribe a well defined quantum state for each atom. This, in turn, makes it difficult to figure out the response of the COSAIN, since our analysis for the COSAIN is based on using the direct product of the quantum state of each atom. For a large value of $N$, it is virtually impossible to develop a manageable density matrix description of the system directly in terms of the collective states. However, it should be possible to evaluate the results of such a density matrix based model for a small value of $N$ ($< 10$, for example). This calculation is a subject of our future work.

For the general case of large $N$, one must rely on an experiment (which, in this context, can be viewed as an analog computer for simulating this problem) to determine the degree of degradation expected from residual spontaneous emission. It should be noted that the detrimental effect of spontaneous emission, for both the CRAIN and the COSAIN, can be suppressed to a large degree by simply increasing the optical detuning while also increasing the laser power. This is the approach used, for example, in reducing the effect of radiation loss of atoms in a far off resonant trap (FORT).

\subsection{\label{subsec:atomfluctuate}Effect of fluctuation in number of atoms}

In both the CRAIN and the COSAIN, the signal is collected multiple times and averaged to increase the signal to noise ratio (SNR). The number of atoms in the ensemble can vary in each run. In the CRAIN, a fluctuation of $\Delta N$ in $N$ is reflected in the signal amplitude by the same amount while the linewidth does not change. This can be easily deduced from the fact that $S_{CRAIN}=N\cos^2(\Delta/2)$. Replacing $N$ by $\Delta N$ will change the signal. However, the FWHM which occurs at $S_{CRAIN}=N/2$, will not change. More details on the classical and quantum noise in the CRAIN and the COSAIN are given in Sec.\ref{subsec:SNR}. In this section, we discuss how the fluctuation in the number of atoms in every run of the experiment affects the signal of the COSAIN.

\begin{figure}[h]
\includegraphics[scale=0.4]{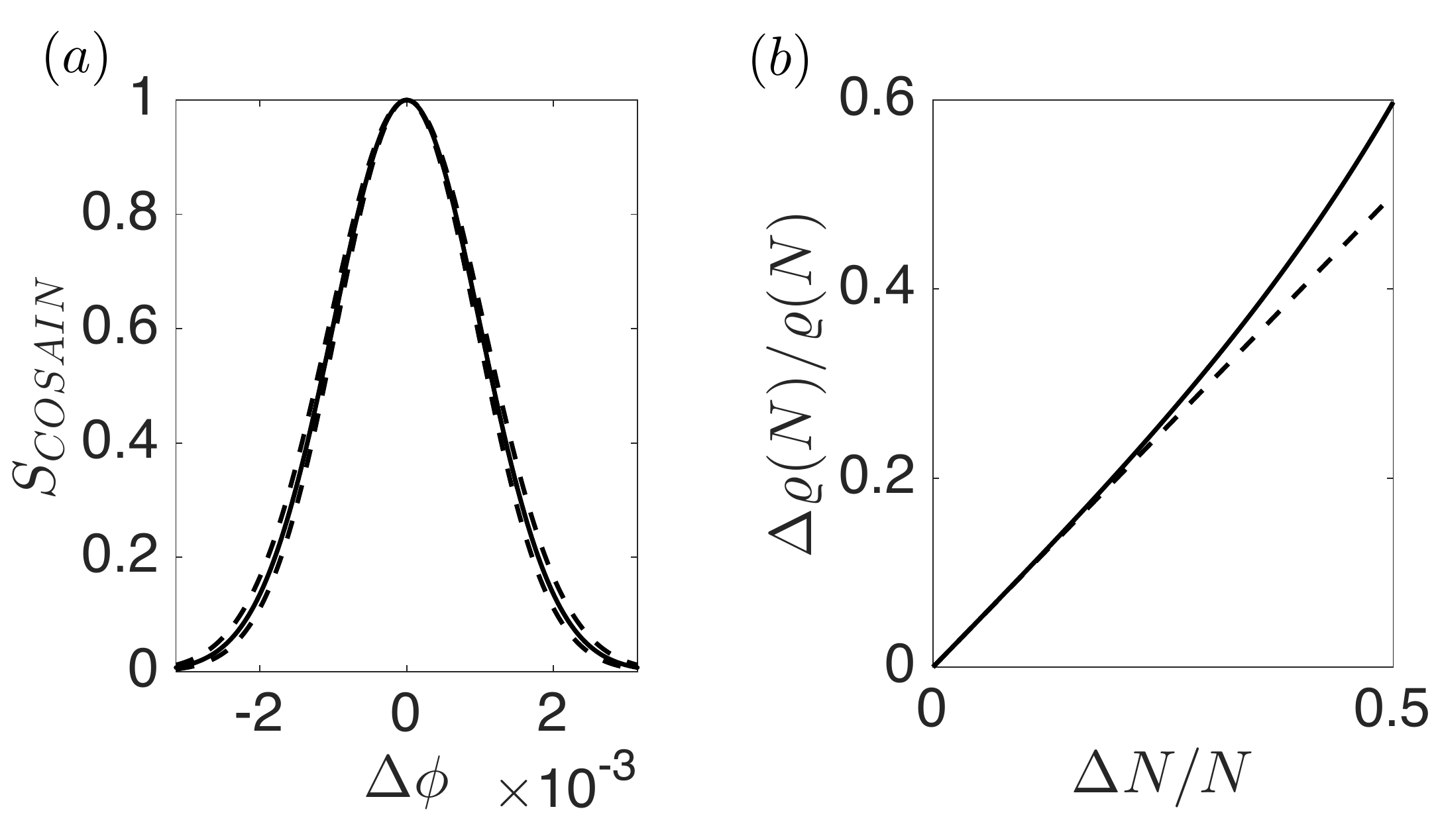}
\caption{\label{Fig_10}(a) $S_{COSAIN}$ for $N=2\times10^5$. (b) Plot of $\Delta\varrho/\varrho$ as a function of $\Delta N/N$.}
\end{figure}

The signal of the COSAIN due to a fluctuation of $\Delta N$ in $N$ is given by $S_{COSAIN}=\cos(\Delta\phi/2)^{2(N\pm\Delta N)}$. Fig.~\ref{Fig_10}(a) shows plot of a COSAIN signal with $N=2\times10^5$. The broken lines represent the case where $\Delta N/N = 0.1$. As is evident from the above discussion, the linewidth increases (decreases) with decreasing (increasing) $\Delta N$. However, the peak of the signal remains at unity, as opposed to the effect of inhomegneity of field and velocity distribution. The signal linewidth of the COSAIN is approximately $\varrho(N)=\varrho(1)/\sqrt{N}$. A fluctuation of $\Delta N$ in $N$ is reflected in the linewidth uncertainty as $\Delta\varrho(N)=\varrho(1)((N-\Delta N)^{-1/2}-(N+\Delta N)^{-1/2})$. The fractional fluctuation is, therefore, $\Delta\varrho(N)/\varrho(N)\simeq(1-\Delta N/N+)^{-1/2}-(1+\Delta N/N+)^{-1/2}=\Delta N/N+0.625(\Delta N/N)^3+0.492(\Delta N/N)^5+O[(\Delta N/N)^6]$. This relation is depicted in Fig.9(b) by the broken line. For small $\Delta N/N$, the fractional change in FWHM is $\Delta\varrho(N)/\varrho(N)\simeq\Delta N/N$ to a good approximation, as shown by the solid line in Fig.~\ref{Fig_10}(b).

\section{\label{sec:experiment}Details of Proposed Experiment}

In order to illustrate the complete picture of the proposed experiment, we consider $^{87}$Rb as the atomic species as an example. We assume a scenario where the atoms will be evaporatively cooled to a temperature of about $2\mu$K, in a dipole force trap~\cite{Olson} and then released. The Raman pulses will be applied while these atoms are falling under gravity. Each Raman pulse will consist of a pair of counterpropagating, right circularly polarized $(\sigma_+)$ beams. One of these beams is red detuned from the $F\!=\!1\!\rightarrow\!F'\!=\!1$ transition in the $D1$ manifold by $\sim1.5$GHz, and the other one is red detuned by the same amount from $F\!=\!2\!\rightarrow\!F'\!=\!1$ transition, also in the $D1$ manifold. The second Raman beam is generated from the first one by a modulator which is driven by an ultrastable frequency synthesizer (FS) tuned to $6.8346826109$ GHz. We assume that the atoms are initially in the $F\!=\!1,\!m_F\!=\!0$ state.

Thus, the states $\ket{g}$ and $\ket{e}$ in Fig.~\ref{Fig_1}(a) would correspond to the hyperfine ground states $F=1,m_F=0$ and $F=2,m_F=0$, respectively. The Raman transitions occur via the excited states $F'=1,m_{F'}=1$ and $F'=2,m_{F'}=1$. The resulting four level system can be reduced to a two level system in the same way as that for the $\Lambda$ system by adiabatically eliminating the excited states together. The resulting system has a coupling rate that is the sum of the two Raman Rabi frequencies. The laser intensities are adjusted to ensure that the light shifts of $\ket{g}$ and $\ket{e}$ are matched.

\begin{figure}
\includegraphics[scale=0.35]{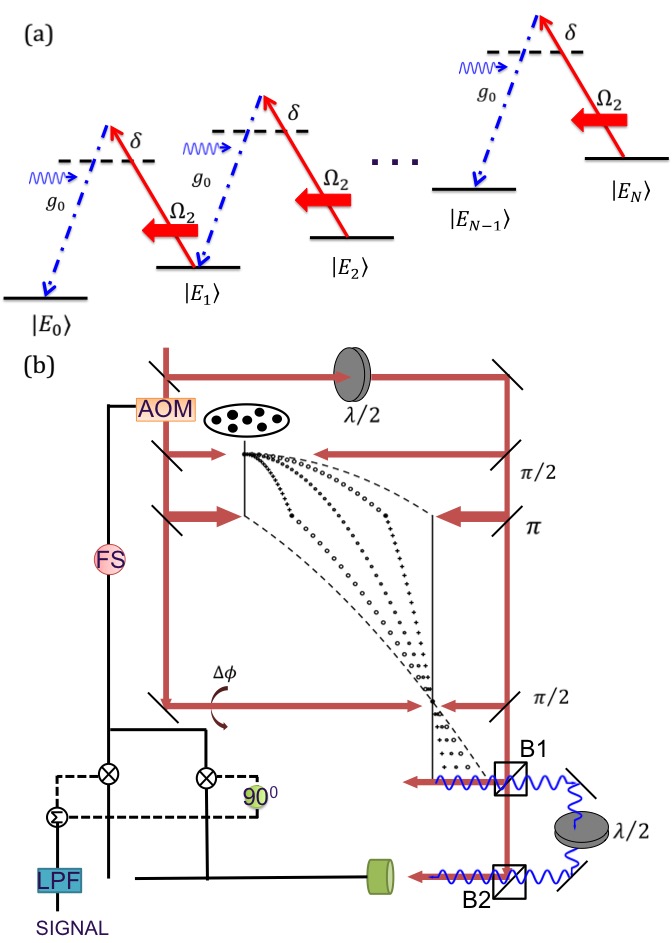}
\caption{\label{Fig_11} (Color online) (a) Interaction between the collective states in the bad cavity limit. (b) Atomic Interferometer experiment for an ensemble of $\Lambda$-type atoms for detecting state $\ket{E_0}$.}
\end{figure}

At the end of the $\pi/2-\pi-\pi/2$ sequence, a probe beam is applied to measure the amplitude of one of the collective states, via the method of zero photon detection. To explain this, we revert to the three-level model of the atom, and first consider a situation where the atomic ensemble is contained in a single mode cavity with volume mode $V$, cavity decay rate $\gamma_c$, resonant at $\omega_1$. The cavity is coupled to the transition $\ket{a}\!\rightarrow\!\ket{g}$ with coupling rate $g_c\!=\!|e.\langle{r}\rangle|E/\hbar$, where $|e.\langle{r}\rangle|$ is the dipole moment of the atom and $E\!=\!\sqrt{2\hbar\omega_1/\epsilon_0V}$. If an off-resonant classical laser pulse of frequency $\omega_2$ is applied, the cavity causes Raman transitions to occur between the collective states $\ket{E_n}$ and $\ket{E_{n-1}}$ with the coupling rates $\Omega'_{n}\!=\!\sqrt{N-n+1}\sqrt{n}\Omega'$, where $\Omega'\!=\!\Omega_2g_c/2\Delta$. This is illustrated in Fig.~\ref{Fig_11}(a).

In the bad cavity limit ($\gamma_c\gg\sqrt{N}\Omega'$), the Raman transitions will still occur. However, the system will not reabsorb the emitted photon, i.e. the transition from $\ket{E_n}$ to $\ket{E_{n-1}}$ will occur, but not vice versa. The field of such a photon is $E\!=\!\sqrt{2\hbar\omega_1/\epsilon_0\mathscr{A}cT}$, where $\mathscr{A}$ is the cross-sectional area of the ensemble and $T$ the interaction time~\cite{DLCZ}. This limit applies here since there is no cavity, so that the stimulated Raman scattering is an irreversible process that can be modeled as a decay with an effective decay rate that is unique to each $\ket{E_n}$. The decay rate from $\ket{E_N}$ is $\gamma_N\!=\!4NL|g_c\Omega_2|^2/\Delta^2c\!=\!N\gamma_{sa}$, where $\gamma_{sa}\!=\!16L\Omega'^2/c$~\cite{Gardiner}, and that for $\ket{E_n}$ is $\gamma_n\!=\!n(N+1-n)\gamma_{sa}$.

 The read beam is extracted from the source and is passed through a $99:1$ ($R/T$: the ratio of the intensity reflectivity, $R$ to the intensity transmittivity, $T$) non-polarizing beam splitter, $B1$ before hitting the ensemble. The probability of counter-propagating photons emitted from this interaction is determined by the resonant optical density of the ensemble.  The direction of signal emission and the role of optical density are discussed further in Sec.~\ref{subsec:OpticalDensity}. The emitted photons pass through $B1$ and, subsequently, through a half-wave plate. The emitted photons and the probe beam are recombined by another $99:1$ beam splitter, $B2$ and sent to a high speed detector, which generates a DC voltage along with a signal at the beat frequency $\sim6.834$ GHz with an unknown phase. This signal is bifurcated and one part is multiplied by the FS signal, while the other is multiplied by the FS signal phase shifted by $90^{\circ}$. The signals are then squared before being combined and sent through a low pass filter (LPF) to derive the DC voltage. This DC voltage is proportional to the number of scattered photons. A lower limit is set for the voltage reading and any values recorded above it will indicate the presence of emitted photons. The duration of the probe beam is set at $\gamma_NT\!=\!10$, where $\gamma_N\!=\!N\gamma_{sa}$ is the slowest decay rate, to ensure that even the longest lived state is allowed to decay almost completely. If no photon is emitted, the voltage will read below the limit, indicating that the ensemble is in state $\ket{E_0}$. If the ensemble is in any other collective state, at least one photon will be emitted. This process is repeated $\mathscr{M}$ times for a given value of $\Delta\phi$. The fraction of events where no photons are detected will correspond to the signal for this value of $\Delta\phi$. This process is then repeated for several values of $\Delta\phi$, producing the signal fringe for a COSAIN. The experimental scheme is illustrated in Fig.~\ref{Fig_11}(b).

\subsection{\label{subsec:OpticalDensity}Role of Optical Density}

In this paper, we have assumed that the ensemble is cigar shaped. This particular choice of configuration is made to achieve the optimum optical density required for realizing the detection scheme discussed above. Consider a four-wave mixing process where three laser beams with wavevectors $\vec{k_1},\vec{k_2},$ and $\vec{k_3}$ interact with a non-linear medium. The process can be viewed as the scattering of the $\vec{k_3}$ beam, for example, off the grating formed by the interference between the $\vec{k_1}$ and $\vec{k_2}$ beams. Efficient phase matching (which is akin to Bragg matching) then requires that the generated beam with a wavevector $\vec{k_4}$ will satisfy the condition that $\vec{k_1}+\vec{k_2}=\vec{k_3}+\vec{k_4}$. The detection process for the COSAIN can be viewed as a time-delayed four wave mixing process. The coherence induced in the ensemble has a spatial variation (i.e. a phase grating) proportional to $\exp(i(\vec{k_1}+\vec{k_2}))$. In the detection zone, we apply a readout field with a wavevector $\vec{k_3}=\vec{k_2}$. Thus, the scattering field will have a wavevector $\vec{k_4}=\vec{k_1}$. This implied that the photon would be scattered in the direction opposite to that of the probe.

In such a scattering process the fraction of photon that would be scattered in directions other than the direction dictated by exact phase matching is determined by the resonant optical density of the ensemble, which is given by $\rho = \sigma n \xi_L$~\cite{DLCZ}. Here, $\sigma\simeq(\lambda/2)^2$ is the resonant scattering cross-section, $n$ is the density, and $\xi_L$ is the interaction length. The fraction of the signal captured by the detector would then be $(\rho-1)/\rho$. This effect can be incorporated in the detector quantum efficiency by writing it as $\eta=\eta_0(\rho-1)\rho$, where $\eta_0$ is the ideal quantum efficiency of the detector.

The proposed detection scheme demands that $\rho\geq 75$, so that at least $98\%$ of the emitted photons are captured, assuming an ideal detector. As discussed in Sec.~\ref{subsec:DopplerShift}, the signal amplitude falls exponentially with increasing ensemble temperature, and $N$. However, the ensemble must not reach the vicinity of critical density at low temperatures. Considering these factors, we choose $N=2.6\times10^4$, $\xi_L=1$mm, and $\xi_T=10\mu$m, deriving $\rho=78.45$ for the $D1$ manifold of $^{87}$Rb.

\section{\label{sec:alternate}Alternate Experimental Scheme}

\begin{figure}[h]
\includegraphics[scale=0.33]{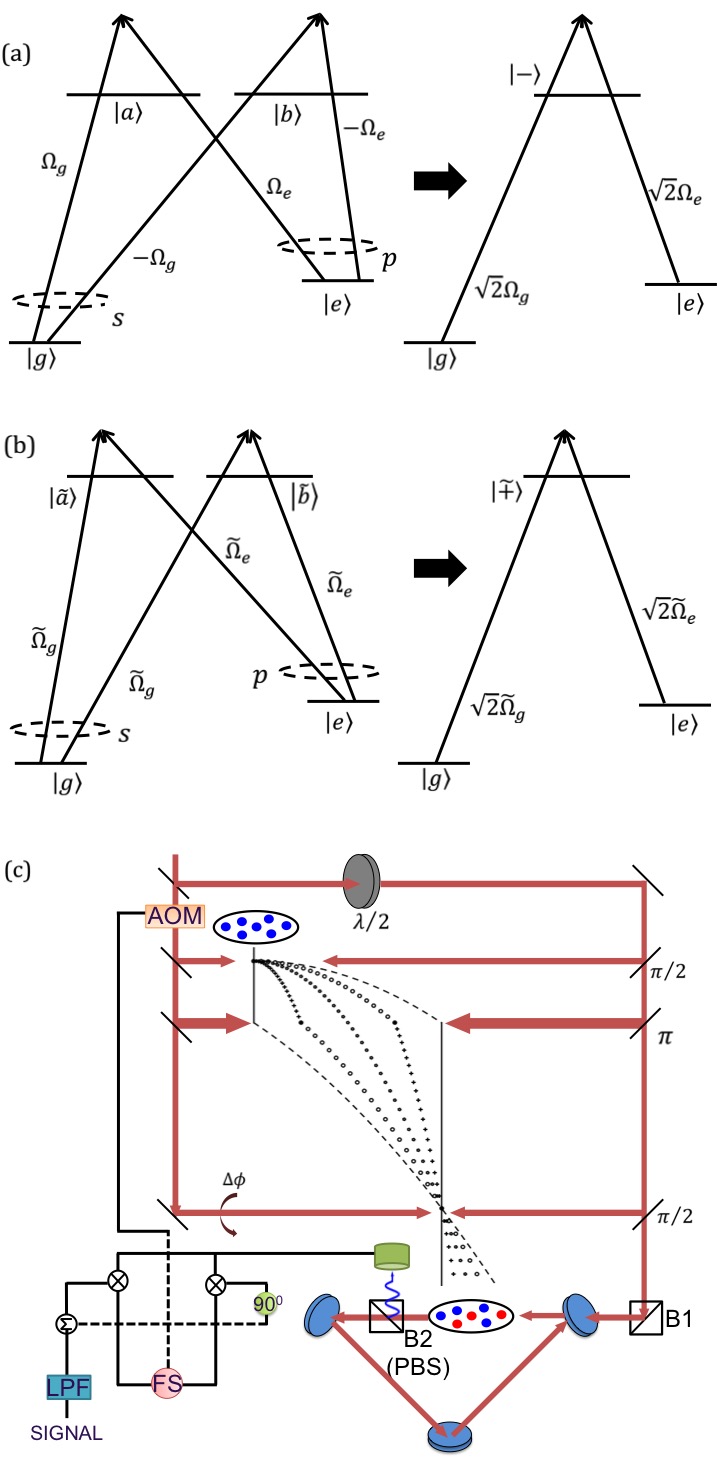}
\caption{\label{Fig_12}(Color online) (a) Raman transitions between $\ket{g\equiv F=1, m_F = 0}$ and $\ket{e\equiv F=2, m_F = 0}$ via $\ket{a\equiv F'=1, m_{F'} = -1}$ and $\ket{b\equiv F'=1, m_{F'} = 1}$, (b) Raman transitions between $\ket{g\equiv F=1, m_F = 0}$ and $\ket{e\equiv F=2, m_F = 0}$ via $\ket{\tilde{a}\equiv F'=2, m_{F'} = -1}$ and $\ket{\tilde{b}\equiv F'=2, m_{F'} = 1}$, (c) Alternate experimental scheme to increase the resonant optical density of the ensemble by introducing a ring cavity in the detection zone.}
\end{figure}

The limitation on cooling the ensemble to reduce the effects of Doppler shift restricts the number of atoms. In turn, this restricts the optical density that can be achieved for an ensemble undergoing the COSAIN sequence. Here, we discuss an alternate experimental scheme that raises the effective optical density of the ensemble. In this scheme, each atom is modeled as a four-level system, as shown in Fig~\ref{Fig_12}(a). The metastable states, $\ket{g}$, and $\ket{e}$, are coupled via two intermediate states, $\ket{a}$, and $\ket{b}$. This four-level system can be reduced to an effective three-level system in the $\Lambda$-configuration. Each Raman pulse will consist of a pair of an $s$-polarized and a $p$-polarized beams, applied in counterpropagating directions. We assume that the $s$-polarized beam is moving in the $+\textbf{z}$-direction, and thus, can be represented as $\mathbf{E}_s=(\hat{\mathbf{\sigma}}^+\tilde{E_{s0}}+\hat{\mathbf{\sigma}}^-\tilde{E_{s0}})\cos(\omega_st-k_sz)=\hat{\mathbf{s}}E_{s0}\cos(\omega_st-k_sz)$. Similarly, the $p$-polarized beam is moving in the $-\textbf{z}$-direction, and thus, can be represented as $\mathbf{E}_p=(\hat{\mathbf{\sigma}}^+\tilde{E_{p0}}-\hat{\mathbf{\sigma}}^-\tilde{E_{p0}})\cos(\omega_pt+k_pz)=\hat{\mathbf{p}}E_{p0}\exp(i\pi/2)\cos(\omega_pt-k_pz)$. Here, $\omega_s$ and $\omega_p$ are the laser frequencies, and $E_{s0}$ and $E_{p0}$ are the amplitudes of the electric field of each laser beam. After making the rotating wave approximation and rotating wave transformation, the atom-laser interaction Hamiltonian elements are $\bra{g}\vec{\rho}\cdot\hat{\mathbf{\sigma}}^+\tilde{E_{s0}}\ket{a}$, $\bra{g}\vec{\rho}\cdot\hat{\mathbf{\sigma}}^-\tilde{E_{s0}}\ket{b}$, $\bra{e}\vec{\rho}\cdot\hat{\mathbf{\sigma}}^+\tilde{E_{p0}}\ket{a}$, $\bra{e}-\vec{\rho}\cdot\hat{\mathbf{\sigma}}^-\tilde{E_{p0}}\ket{b}$, and the corresponding complex conjugates. Here $\vec{\rho}=x\hat{\mathbf{x}}+y\hat{\mathbf{y}}+z\hat{\mathbf{z}}=\rho_{\sigma^+}\hat{\mathbf{\sigma^+}}+\rho_{\sigma^-}\hat{\mathbf{\sigma^-}}+z\hat{\mathbf{z}}$. The Hamiltonian can be further simplified to $H=\rho_{ga}\tilde{E}_{s0}\ket{g}\bra{a}+\rho_{gb}\tilde{E}_{s0}\ket{g}\bra{b}+\rho_{ea}\tilde{E}_{p0}\ket{e}\bra{a}-\rho_{eb}\tilde{E}_{p0}\ket{e}\bra{b}+c.c.$

For concreteness, we use the $D1$ line of $^{87}$Rb to illustrate the mechanism behind this scheme. Thus, the states $\ket{g}$ and $\ket{e}$ in the left part of Fig.~\ref{Fig_12}(a) would correspond to the hyperfine ground states $F = 1,m_F = 0$ and $F = 2, m_F = 0$, respectively. The Raman transitions occur via the excited states $\ket{a\equiv F' = 1,m_{F'}= -1}$ and $\ket{b\equiv F' = 1, m_{F'}= 1}$. For this particular choice of levels, $\rho_{ga}=-\rho_{gb}=\rho_{ea}=\rho_{eb}=|\rho_0|$, $|\rho_0\tilde{E_{s0}}|=\hbar\Omega_g/2$, and $|\rho_0\tilde{E_{p0}}|=\hbar\Omega_e/2$. The atom-laser interaction Hamiltonian in this case is, therefore, $H=\hbar(\Omega_{g}\ket{g}\bra{a}-\Omega_{g}\ket{g}\bra{b}+\Omega_{e}\ket{e}\bra{a}-\Omega_{e}\ket{e}\bra{b})/2+c.c.$ This four level system can be reduced to an equivalent three-level model by rotating the $\{\ket{a},\ket{b} \}$ Hilbert sub-space by $\pi/4$. The reduced Hamiltonian, $H_{red}$ is given by $H_{red}=\hbar(\Omega_g\ket{g}\bra{-}+\Omega_e\ket{e}\bra{-})/\sqrt{2}+c.c.$, where $\ket{-}=(\ket{a}-\ket{b})/\sqrt{2}$, as illustrated in the right part of Fig.~\ref{Fig_12}(a). The $D1$ line of $^{87}$Rb are also coupled via $\ket{\tilde{a}\equiv F' = 2,m_{F'}= -1}$ and $\ket{\tilde{b}\equiv F' = 2, m_{F'}= 1}$. In this case, $-\rho_{g\tilde{a}}=-\rho_{g\tilde{b}}=-\rho_{e\tilde{a}}=\rho_{e\tilde{b}}=|\tilde{\rho_0}|$, $|\tilde{\rho_0}\tilde{E_{s0}}|=\hbar\tilde{\Omega_g}/2$, and $|\tilde{\rho_0}\tilde{E_{p0}}|=\hbar\tilde{\Omega_e}/2$. Thus, the Hamiltonian is $\tilde{H}=-\hbar(\tilde{\Omega_g}\ket{g}\bra{\tilde{a}}+\tilde{\Omega_g}\ket{g}\bra{\tilde{b}}+\tilde{\Omega_e}\ket{e}\bra{\tilde{a}}+\tilde{\Omega_e}\ket{e}\bra{\tilde{b}})/2+c.c.$ The reduced equivalent three-level Hamiltonian is $\tilde{H}_{red}=\hbar(\tilde{\Omega_g}\ket{g}\bra{\tilde{+}}+\tilde{\Omega_e}\ket{e}\bra{\tilde{+}})/\sqrt{2}+c.c.$, where $\ket{\tilde{+}}=(\ket{\tilde{a}}+\ket{\tilde{b}})/\sqrt{2}$. These transitions are shown in the Fig.~\ref{Fig_12}(b). Thus, the system is equivalent to two $\Lambda$-systems, each with a different common mode detuning.  Adiabatic elimination of the $\ket{-}$ and the $\ket{\tilde{+}}$ states would produce the effective two level transition between $\ket{g}$ and $\ket{e}$, just as in the case of excitations with circularly polarized fields described earlier.

At the end of the $\pi/2$-dark-$\pi$-dark-$\pi/2$ sequence, the ensemble is introduced into a ring cavity of finesse $\mathscr{F}$. The read beam is extracted from the $p$-polarized beam and enters the cavity through port $P1$, as illustrated in Fig.~\ref{Fig_12}(c). The scattered photons, which will be $s$-polarized, are extracted with a polarizing beam splitter, B2.  Note that this type of extraction is not possible if the interferometer were to be realized with circularly polarized beams. The repeated interaction of the ensemble with the read beam increases the effective resonant optical density of the ensemble to $\mathscr{F}\rho/\pi$. Since the ensemble is falling under gravity through the course of the experiment, the cavity mode size must be reasonably large to accommodate this motion. We assume that the length of the first dark zone is $1$cm, and that the distance between the last $\pi/2$ pulse and the read beam is also $1$cm. The duration of the read beam, $T$ is set at $\gamma_NT =10$, where $\gamma_N =N\gamma_{sa}$ is the slowest decay rate, to ensure that even the longest lived state is allowed to decay completely. It can be shown that for $N=2.6\times 10^4$, $T\simeq 3.3$ms, so that the distance traveled by the ensemble during the interrogation period is $\simeq 3.3$mm. The cavity mode size must be at least twice as much as this distance.

\section{\label{sec:comparison}Performance of the COSAIN Compared to that of the CRAIN}

In order to compare the performance of the COSAIN to that of the CRAIN, we analyze the stability of the phase-difference measured by them by investigating the fluctuation that has both quantum mechanical and classical components, i.e. $\delta\Delta\phi|_{total}\!=\!(\Delta S_{QM}\!+\!\Delta S_{classical})/|\partial\!S/\partial\!\Delta\phi|$, where $S(\Delta\phi)$ is the signal. Since the signal depends on the phase, the fluctuation is not necessarily constant. Therefore, there is no unique value of signal to noise ratio (SNR) to compare unless the COSAIN and the CRAIN are compared at a particular value of the phase-difference. Thus, the fluctuations must be compared as a function of $\Delta\phi$. In Sec.~\ref{subsec:SNR}, we discuss in detail, the quantum fluctuation due to quantum projection noise, $\Delta\!P\!=\!\sqrt{P(1-P)}$~\cite{Wineland}, where $P$ is the population of the state being measured, and the classical noise in the long term regime. Since the measure of the signal depends on counting zero photon events, the efficiency of the high speed detector affects the signal amplitude and width. In Sec.~\ref{subsec:detector}, we discuss the effect of the detector efficiency on the COSAIN signal. In Sec.~\ref{subsec:CollEff}, we discuss the collection efficiency of the COSAIN as a measure of its performance as compared to the CRAIN. The CRAIN suffers from imperfect collection efficiency due to the latter's dependence on experimental geometry. On the other hand, the collection efficiency of the COSAIN is close to unity owing to the fact that the fluorescence of photons is collected through coherent Raman scattering. As a result, for the same number of atoms detected per unit time, the COSAIN is expected to outperform the CRAIN by as much as a factor of $10$.

\subsection{\label{subsec:SNR}Effect of Quantum and Classical Noise}

For the COSAIN to be a useful device for practical metrology, it must outperform the CRAIN. To explore this, we compare their stability in the short term and the long term regimes. The stability of an interferometer is determined by the fluctuations in $\Delta\phi$ that has both quantum mechanical and classical components. The phase difference, $\Delta\phi$ (expressed in radians) is proportional to the rate of rotation of the gyroscope, $\Omega_G$ (see Sec.~\ref{sec:narrowing}). Thus, $\Delta\phi=\mu\Omega_G$, where $\mu$ depends on the area of the interferometer and mass of the single atom.

In the CRAIN described above, the signal is a measure of the probability of finding the atom in state $\ket{g}$, $P_g=\cos^2(\mu\Omega_G/2)$. The signal is detected by probing the desired state for a duration of time. If $\tilde{N}$ is the number of atoms per unit time and $T$ is the interrogation period, then the net signal is $S_{CRAIN}=\tilde{N}TP_g$. For comparison, we set the number of atoms per trial in the COSAIN, $N$, multiplied by the number of trials, $M$, to equal $\tilde{N}T$. Therefore, $S_{CRAIN}=MN\cos^2(\mu\Omega_G/2)$. Since the fluctuation in $MN$ is $\sqrt{MN}$, the quantum mechanical variance of the signal is $\Delta(S_{CRAIN,QM})=\sqrt{MN}\sin(\mu\Omega_G)/2$, since the projection noise in a single two level atomic system is $\Delta S_{CRAIN}=\sqrt{P_g(1-P_g)}$~\cite{Wineland}. In the case where the probability of finding the atom in $\ket{g}$ is $0$ or $1$, the projection noise vanishes. On the other hand, the projection noise is at its peak value when $P_g=1/2$. The slope of the signal is, therefore, $\partial S_{CRAIN}/\partial\Omega_G=-MN\sin(\mu\Omega_G)/(2\gamma_{sa})$, where $\gamma_{sa}=1/2\mu$ is the linewidth. Assuming ideal quantum efficiency of the detection process, the fluctuation in the rate of rotation can be written as $\delta\Omega_G|_{total}=|(\Delta S_{QM}+\Delta S_{classical})/(\partial S_{CRAIN}/\partial\Omega_G)|$, which maybe be considered as noise $(\Delta S)$, over the Rotational Variation of Signal (RVS) which is $(\partial S_{CRAIN}/\partial\Omega_G)$. In the following text, we consider first the effect of quantum noise. The quantum rotation-rate fluctuation (QRF) for a CRAIN maybe written as 
\begin{align}
\delta\Omega_G|_{QM,CRAIN}=\left\lvert\frac{\Delta S_{QM}}{(\partial S_{CRAIN}/\partial\Omega_G)}\right\lvert=\frac{\gamma_{sa}}{\sqrt{MN}}.
\label{Eq_QRF_CRAIN}
\end{align}
It is, thus, merely a coincidence that the QRF turns out to be constant in a CRAIN. Contrary to popular perception, the QRF of an interferometer is, therefore, not fundamentally the linewidth divided by the SNR. It should be evident from the above discussion that the signal is not given by $MN$, and the noise is not given by $\sqrt{MN}$. Instead, they both depend on $\Omega_G$.
 
In devices where the QRF is not a constant, as we will show for a COSAIN, it is thus, imperative that we carry out an analysis of the QRF in a manner analogous to the analysis for the CRAIN shown above. Thus, we will adopt the approach that the net rotation-rate fluctuation, $\delta\Omega_G$ should be thought of as the ratio of the noise to the RVS. This approach should be adopted universally for all metrological devices. Of course, for devices where the relevant quantity is not the rotation rate, the definition should be adapted accordingly. For example, in a clock that measures frequency, the relevant quantity can be expressed as follows: net frequency fluctuation is the ratio of the noise to the Spectral Variation of Signal (SVS).

\begin{figure}[h]
\centering
\includegraphics[scale=0.33]{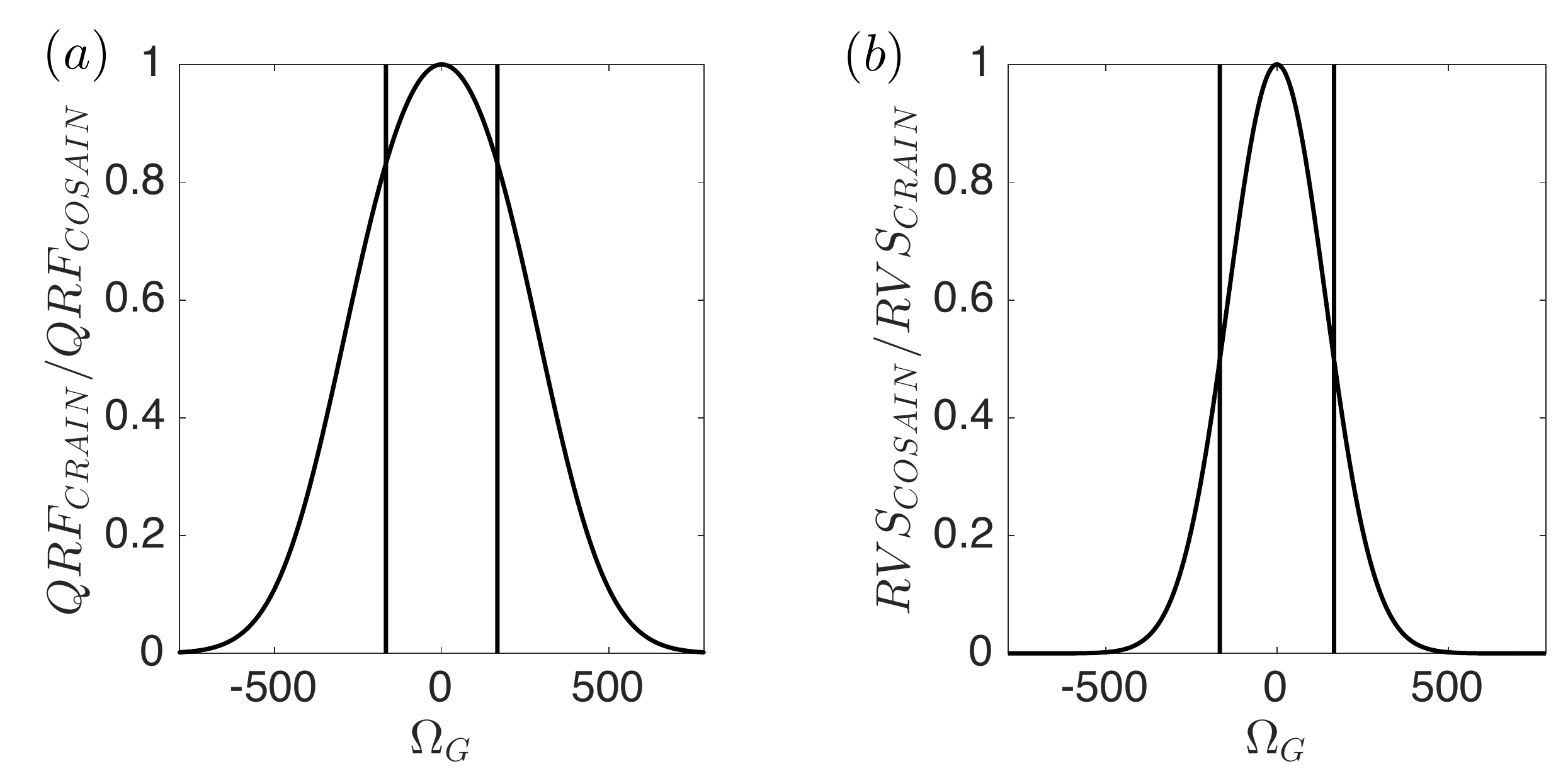}
\caption{\label{Fig_13}(left) Ratio of the QRF in the CRAIN to the QRF in the COSAIN, for $M = 1000$ and $N = 10^4$. It should be noted that the fluctuation in the CRAIN is independent of $\Omega_G$ while that of the COSAIN varies significantly with it. (right) Ratio of the RVS of the COSAIN to the RVS of the CRAIN for $M = 1000$ and $N = 10^4$. The vertical lines in the plots show where the FWHM of $S_{CRAIN}$ are.}
\end{figure}

Following this approach, we calculate the net rotation-rate fluctuation of the COSAIN and compare it to that of the CRAIN. We will first calculate the quantum fluctuation which is relevant in the short term regime, and then the classical fluctuation, which dominates in the long term regime. The signal of a COSAIN for $M$ trials is $S_{COSAIN}=MP_{E0}=M\cos^{2N}(\mu\Omega_G/2)$, and the projection noise is $\Delta P_{E0}=\sqrt{P_{E0}(1-P_{E0})}$ for a single trial, so that $\Delta P_{E0}=\sqrt{M}\sqrt{P_{E0}(1-P_{E0})}$ for $M$ trials. Thus, the total quantum mechanical noise in the signal is 
\begin{align}
 \Delta P_{E0}=\sqrt{M}\cos^{N}(\mu\Omega_G/2)\sqrt{1-\cos^{2N}(\mu\Omega_G/2)},
\end{align}
and the RVS is
\begin{align}
\partial S_{COSAIN}/\partial\Omega_G &=-MN\cos^{2N-1}(\mu\Omega_G/2)\nonumber\\
&\times\sin(\mu\Omega_G/2)/\gamma_{sa}.
\end{align}
Therefore, the QRF in the COSAIN is given by
\begin{align}
\delta\Omega_{G(QM,COSAIN)}=\frac{\gamma_{sa}}{N\sqrt{M}}\frac{\sqrt{\sec^{2N}(\mu\Omega_G/2)-1}}{\tan(\mu\Omega_G/2)}.
\label{Eq_QRF_COSAIN}
\end{align}
Thus, unlike the CRAIN, the phase fluctuation in a COSAIN is not constant and depends on $\Omega_G$ and thus, on $\Delta\phi$. We consider first the limiting case of $\Omega_G\rightarrow 0$. Using Taylor expansion, it is evident that $\delta\Omega_{G(QM,COSAIN)}=\gamma_{sa}/\sqrt{MN}$, which is the same as that of a CRAIN. This can be understood physically by noting that while the fringe width becomes much narrower for the COSAIN, the SNR also decreases due to the fact that a single observation is made for all N atoms in a given trial. The QRF for the COSAIN, given in Eq.~\ref{Eq_QRF_COSAIN}, is smallest as $\Omega_G\rightarrow 0$ and increases as $\Omega_G$ moves
away from zero. The ratio of the QRF for the CRAIN to that of the
COSAIN is plotted as a function of $\Omega_G$ in the left side of Fig.~\ref{Fig_13} for $M = 1000$ and $N = 10^4$. Here, the vertical bars indicate the FWHM of the COSAIN signal. It is clear from this plot that the QRF for the COSAIN increases significantly as we move away from resonance. However, since a servo will keep the value of $\Omega_G$ confined to be close to zero, the phase stability of the COSAIN, under quantum noise limited operation, should be very close to that of the CRAIN, assuming that all the other factors remain the same.

The classical rotation-rate fluctuation (CRF), $\delta\Omega_G|_{classical} = \Delta S_{classical}/(\partial S/\partial\Omega_G)$, is the limiting factor in the long term stability. While the quantum fluctuation is dominated by quantum projection noise, the classical noise is dominated by noise in the electronic and the mechanical components employed to generate the interferometer signal. Since the pieces of equipment used in the development of both the COSAIN and CRAIN suffer from similar noise issues, the variance $\Delta S$ is expected to be of the same order of
magnitude for both interferometers. On the other hand, the RVS, $(\partial S/\partial\Omega_G)$, is not the same, as was shown previously. The ratio of the RVS of the COSAIN to the RVS of the CRAIN is
\begin{align}
\frac{\partial S_{COSAIN}/\partial\Omega_G}{\partial S_{CRAIN}/\partial\Omega_G}=\frac{cos^{2N}(\mu\Omega_G/2)}{cos^{2}(\mu\Omega_G/2)}=\frac{P_{E0}}{P_g},
\end{align}
and is plotted in Fig.~\ref{Fig_8}(right). With $\Delta S_{classical,COSAIN}\sim\Delta S_{classical,CRAIN}$, the ratio of the CRF of the COSAIN to that of the CRAIN can be written as
\begin{align}
\frac{\delta\Omega_{G(classical,COSAIN)}}{\delta\Omega_{G(classical,CRAIN)}}\simeq\frac{\cos^{2}(\mu\Omega_G/2)}{\cos^{2N}(\mu\Omega_G/2)}.
\label{Eq_ratio_CRF}
\end{align}

Similar to the ratio of the two interferometers in QRF, Eq.~\ref{Eq_ratio_CRF} is smallest as $\Delta\phi\rightarrow 0$ and increases as $\Delta\phi$ moves away from resonance. Thus, with respect to both quantum and classical sources of noise, the COSAIN must be operated near $\Delta\phi\simeq 0$ for optimal performance.

\subsection{\label{subsec:detector}Effect of Detector Efficiency}

The key aspect of the COSAIN is the measurement of the amplitude of $\ket{E_0}$, which indicates that each of the atoms in the ensemble is individually in $\ket{g}$. The probe beam is applied to the ensemble, which is in the quantum state $\ket{\Psi}=c_0\ket{E_0}+\sum_{n=1}^N\ket{E_n}$. Interaction between the probe beam, the ensemble, and the free space vacuum modes on the other leg would lead to production of photons unless $c_0=1$, and $c_n=0$ for all $n$. These photons are detected using a heterodyning technique, as described in  Sec~\ref{sec:description}. The voltage output of the heterodyning system is proportional to the amplitude of the electric field corresponding to the photons.

In general, one or more photons are produced as $\ket{E_n}$ decays to $\ket{E_{n-1}}$ and subsequent states. The time needed for these photons to be produced depends in the vacuum and probe field induced Raman transition rates between $\ket{E_n}$ and $\ket{E_{n-1}}$. If we assume perfect efficiency for detecting each of these photons, and wait for a time long compared to the inverse of the weakest of these transition rates, then the detection of no photons implies that the system is in state $\ket{E_0}$. In practical experimental conditions, we can choose a small threshold voltage at the output of the heterodyning system as an indicator of null detection. Thus, any signal below this threshold would be viewed as detection of the quantum system in the $\ket{E_0}$ state, and all signals above this threshold would be discarded. The number of events below this threshold for $\mathscr{M}$ trials carried out with all the parameters of the experiment unchanged, is the derived signal for the COSAIN. After collecting data for all the values of $\Delta\phi$ that is of interest, the result would ideally yield the plot of the COSAIN signal $S_{COSAIN}=|c_0|^2$, averaged over $\mathscr{M}$ trials. However, with a fractional detector efficiency and finite detection period, the signal would deviate from the ideal result.

Consider first the effect of the detection period. Given the decay rate of the off-resonant Raman process, $\gamma_n=n(N+1-n)\gamma_{sa}$, the probability that $\ket{E_n}$ will produce zero photons during the measurement period $T$ is $P_{0,n}=e^{-\gamma_nT}$. Thus, the total probability of zero photon emission (which should vanish ideally for any $c_n\neq 0$) is given by $P_0=\sum_{n=1}^{N}|c_n|^2e^{-\gamma_nT}$. The COSAIN signal, $S_{COSAIN}$, is the total probability of finding zero photons during $T$, and can be expressed as $S_{COSAIN}=|c_0|^2+\sum_{n=1}^{N}|c_n|^2e^{-\gamma_nT}$. Noting that $\gamma_0=0$, we can rewrite this as $S_{COSAIN}=\sum_{n=0}^{N}|c_n|^2e^{-\gamma_nT}$. The lower and upper bounds of $S_{COSAIN}$ can be established by considering the strongest and the weakest effective decay rates. The strongest decay rate occurs for the middle state, $\gamma_{N/2}=(N/2)(N/2+1)\gamma_{sa}\approx(N^2/4)\gamma_{sa}$, where $N\gg 1$ approximation has been made. With the substitution of the largest decay rate for each $\ket{E_n}$ into the equation for $S_{COSAIN}$, the lower bound is set by
\begin{align}
P_L=|c_0 |^2+(1-|c_0 |^2) \exp({-N^2\gamma_{sa}T/4}).
\label{Eq_COSAIN_PL}
\end{align}
On the other hand, the weakest decay rate is exhibited when $n=N$, making the upper bound on the signal,
\begin{equation}
P_U=|c_0|^2+(1-|c_0 |^2)\exp({-N\gamma_{sa}T}). 
\label{Eq_COSAIN_PU}
\end{equation}
The signal of the COSAIN, $S_{COSAIN}$ produced in time $T$ will lie somewhere between $P_L$ and $P_U$.

Consider next the effect of the non-ideal detection efficiency of the heterodyning scheme. For concreteness, we define $\eta$ as the efficiency of detecting a single photon. In practice, this parameter will depend on a combination of factors, including the quantum efficiency of the high-speed photon detector and the overlap between the probe laser mode and the mode of the emitted photon. For the present experiment, we are only interested in knowing whether at least one photon is detected, and not in the actual number of photons. When more photons are emitted, the detector will have a better chance of observing a non-zero signal, and hence distinguish dark counts from the rest with more certainty. For example, if three photons are emitted in time $T$, then four different outcomes are possible:
\begin{itemize}
\item All three photons are detected, with probability $\eta^3$;
\item Two of the photons are detected with probability $\eta^2(1-\eta)$; this can occur for any two of the photons, so the multiplicity is $3$;
\item One photon is detected, with probability $\eta(1-\eta)^2$ and multiplicity of $3$;
\item No photons are detected, with probability $\epsilon^3\equiv(1-\eta)^3$.
\end{itemize}
The sum of these probabilities is $1$. The probability that at least $1$ photon is detected is thus $(1-\epsilon^3)$. For any state $n\neq 0$, the probability of detecting at least $1$ photon is, therefore, $(1-\epsilon^{n})$.

Moreover, we must also consider how the effective detection efficiency is influenced by the fact that the collective states decay at different rates. Specifically, the $n$ level for $n>0$ might produce $N-n$ photons, $N-n-1$ photons, down to zero photons, depending on the length of the measurement time and the effective decay rate. If the system is in $\ket{E_{3}}$, for example, it can produce up to $3$ photons but with probabilities that change over $T$. For a given time $T$, $\ket{E_3}$ evolves into a sum of the states $\ket{E_3}\rightarrow\sum_{k=0}^{3}a_{n,k}(T)\ket{E_k}$, where the coefficient $a_{n,k}(T)$ depends on the effective decay rate that is specific to each state, and changes as the states evolve in time. Thus the probability of detecting at least one photon is
\begin{align}
P=\sum_{n=1}^{N-1}|c_n|^2\sum_{k=n}^{N}(1-\epsilon^{k-n})|a_{n,k}(T)|^2.
\end{align}
Therefore, the probability of detecting no photon is
\begin{align}
S_{COSAIN}=1-P &= 1-\sum_{n=1}^{N-1}|c_n|^2 \nonumber\\
&\times\sum_{k=n}^{N}(1-\epsilon^{k-n})|a_{n,k}(T)|^2.
\end{align}

The numerical analysis for a large number of atoms is tedious and scales as at least $(N-1)!$ for the COSAIN. However, we can take the worst case scenario to serve as the upper bound for the signal. The worst case occurs when only a single photon is produced as a result of $\ket{E_n}$ decaying to only the $\ket{E_{n-1}}$ state, so that the index of the second summation stops at $k=n-1$. In this case, we can write $|a_{n,n-1}(T)|=(1-e^{-\gamma_nT})$ and the signal becomes
\begin{align}
S_{COSAIN}=|c_0|^2+\epsilon(1-|c_0|^2)+\eta\sum_{n=1}^{N}|c_n|^2e^{-\gamma_nT}.
\end{align}
Using the approach we employed in arriving at Eq.~\ref{Eq_COSAIN_PL} and Eq.~\ref{Eq_COSAIN_PU}, we now consider the strongest and the weakest decay rates for single photon production to arrive at the lower and upper bounds of the zero photon count signal:
\begin{align}
P_L = 1-\eta(1-|c_0|^2)(1-e^{-N^2\gamma_{sa}T/4}),\\
P_U = 1-\eta(1-|c_0|^2)(1-e^{-N\gamma_{sa}T}).
\end{align}

\begin{figure}[h]
\includegraphics[scale=0.35]{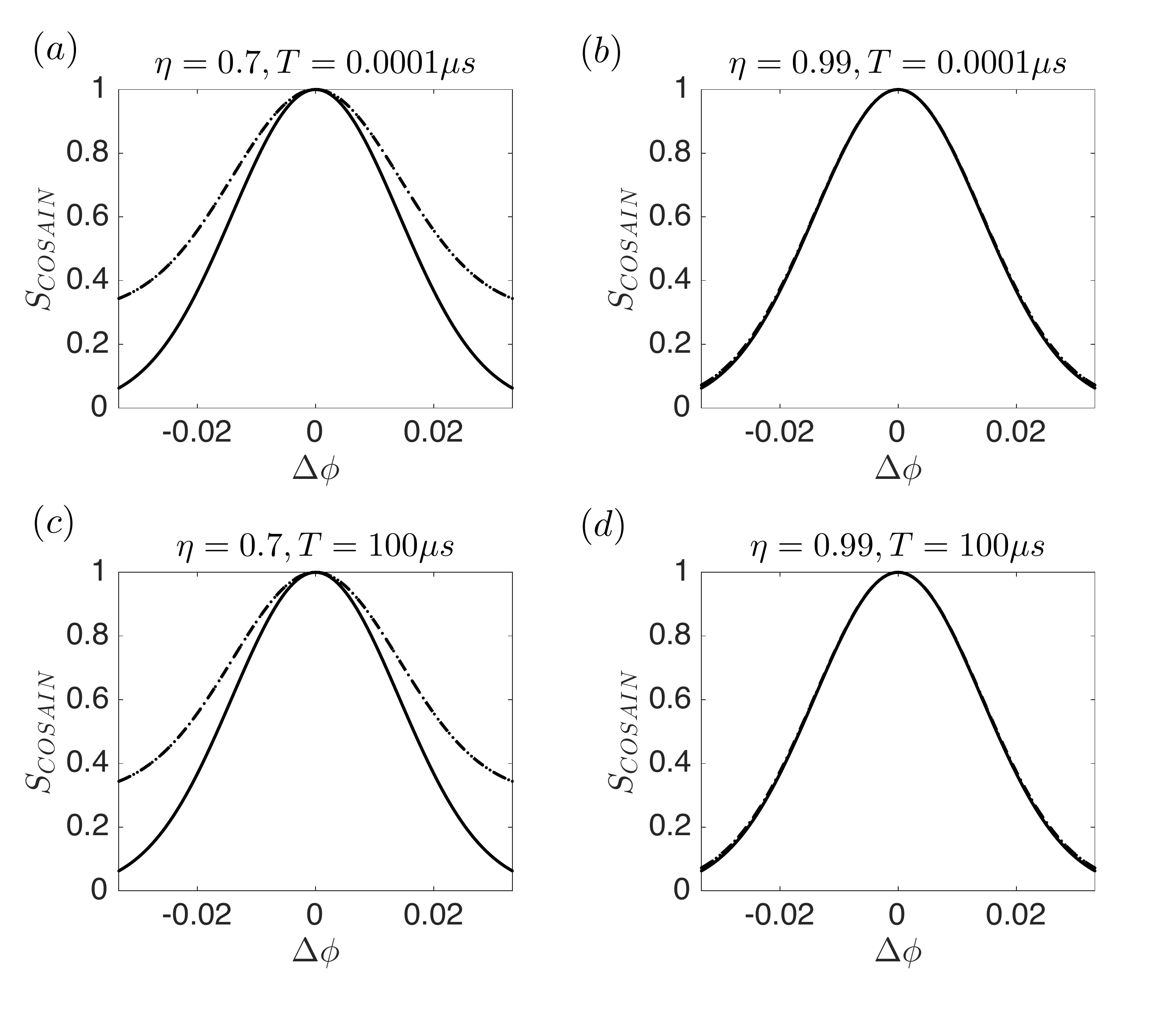}
\caption{\label{Fig_14} Plot of ideal signal (solid line), the upper bound (broken line), the lower bound (dotted line) for different detection times, $T$ and detector efficiencies, $\eta$ for $N=10,000$.}
\end{figure}

Figure~\ref{Fig_14} shows the plot of the ideal $S_{COSAIN}, P_L$ and $P_U$ over a variation in $\Delta\phi$ for different values of detector efficiencies and detection times for $N=10,000$. It can be seen from the plots that the upper and lower bounds on the signal coincide with the ideal signal in the vicinity of $\Delta\phi\rightarrow 0$. For a larger size of the ensemble, a longer detection time ensures that the gap between the bounds decreases and that they are closer to the ideal signal.

If we set $\gamma_{sa}T=1$, the signal depends on $\eta$ as
\begin{align}
S_{COSAIN}\simeq 1-\eta(1-\cos^{2N}(\Delta\phi/2))
\label{Eq_COSAIN_Signal_nonideal}
\end{align}
for large $N$ and $\mathscr{M}=1$. Hence, we can calculate the QRF for the COSAIN to see how it depends on the detector efficiency, and how it compares to the CRAIN. For the CRAIN, it is straightforward to show that with $S_{CRAIN}=\eta N\cos^2(\Delta\phi/2)$, the quantum mechanical noise in the signal is $\Delta S_{CRAIN}=\sqrt{\eta N}\cos(\Delta\phi/2)\sin(\Delta\phi/2)$ and the RVS is $|\partial S_{CRAIN}/\partial\Omega_G|=(\eta N/\gamma_{sa})\cos(\Delta\phi/2)\sin(\Delta\phi/2)$, so that the QRF is $\delta\Omega_{G(QM,CRAIN)}=\gamma_{sa}/\sqrt{\eta N}$. It is also straightforward to calculate the QRF of the COSAIN. the total quantum mechanical noise in the COSAIN signal in Eq.~\ref{Eq_COSAIN_Signal_nonideal} is:
\begin{align}
\Delta S_{QM,COSAIN}=\eta\cos^N(\Delta\phi/2)\sqrt{1-\cos^{2N}(\Delta\phi/2)},
\end{align} 
and the RVS is
\begin{align}
\partial S_{COSAIN}/\partial\Omega_G = -(\eta N/\gamma_{sa})\sin(\Delta\phi/2)\cos^{2N-1}(\Delta\phi/2).
\end{align}
Thus, the QRF of the COSAIN is
\begin{align}
\delta\Omega_{G(QM,COSAIN)}=\left\lvert\frac{\gamma_{sa}}{N\sqrt{\eta}}\frac{\sqrt{1-\cos^{2N}(\mu\Omega_G/2)}}{\cos^{N-1}(\mu\Omega_G/2)\sin(\mu\Omega_G/2)}\right\rvert
\end{align}
which approaches $\gamma_{sa}/\sqrt{\eta N}$ as $\Omega_G\rightarrow 0$. Assuming that the detector efficiencies of the COSAIN and the CRAIN can be essentially the same, they do not affect the ratio of the two QRF's.

\subsection{\label{subsec:CollEff}Effect of Collection Efficiency}

We consider next the effect of collection efficiency, $\beta$ on the COSAIN and compare it to that of the CRAIN. The signal for both the COSAIN and the CRAIN, is directly proportional to $\beta$. From Eq.~\ref{Eq_QRF_CRAIN} and Eq.~\ref{Eq_QRF_COSAIN}, it is easy to show that 
\begin{align}
\zeta &\equiv\frac{\delta \Omega_{G(QM,COSAIN)}}{\delta \Omega_{G(QM,CRAIN)}} \nonumber\\
 &=\frac{\sqrt{\sec^{2N}(\frac{\mu\Omega_G}{2})-1}}{\sqrt{N}\tan(\frac{\mu\Omega_G}{2})}\sqrt{\frac{\beta_{CRAIN}}{\beta_{COSAIN}}},
\label{Eq_Collection_Efficiency}
\end{align}
where $\beta_{CRAIN}\space(\beta_{COSAIN})$ is the collection efficiency of the CRAIN (COSAIN).

As $\Omega_G\rightarrow 0$, the quantity in the square bracket in Eq.~(\ref{Eq_Collection_Efficiency}) approaches unity. Therefore, in this limit, $\zeta$, the ratio of the QRF of the COSAIN to that of the CRAIN, would depend on the ratio of of the collection efficiencies of the detection process. The coherent stimulated Raman scattering based detection method used for the COSAIN process has a collection efficiency that is close to unity, i.e. $\beta_{COSAIN}\simeq 1$. In the case of the CRAIN, the fluorescence is collected from the spontaneous emission process, which emits photons in a dipolar radiation pattern. The $\beta_{CRAIN}$ can be quantified by analyzing the detection method, for example, of a CRAIN that makes use of cold atoms released from a MOT. For a lens placed at a distance of $r = 5$ cm, with a diameter of $d = 2.5$ cm, ignoring the dipolar pattern of radiation for simplicity, and assuming it to be uniform in all directions, this system yields a value of $\beta_{CRAIN}\simeq d^2/4r^2= 1/16$ corresponding to $\zeta \sim 0.25$. In a typical CRAIN, various geometric constraints make it difficult to achieve a value of $\beta_{CRAIN}$ much larger than this. In practice, in cases where the total volume occupied by the CRAIN has to be constrained in order to meet the user requirements, the value of $\beta_{CRAIN}$ is typically $1\%$, which would correspond to $\zeta \sim 0.1$. Thus, the near unity collection efficiency of the COSAIN can lead to an improvement of the interferometer stability by as much as a factor of 10.

Another method of detecting signal in a CRAIN is absorption. However, the use of absorption warrants the consideration of many practical issues. The fluctuation in $\Delta\phi$ is affected by additional noise contributed by the laser used in absorption. Let us assume that the observation time is $T$, and the number of photon in the probe beam before absorption is $N_P$, and that the probe is in a Coherent state. Furthermore, we assume that the number of atoms passing through the detection process within this time is $N_A$, and the linewidth of resonance is $\Gamma$. If the detection process produces an absorption by a fraction of $\kappa$ (i.e. $\kappa=1$ represents  perfect absorption of the laser beam), and the detector has a quantum efficiency of $\eta$, then the resulting fluctuation in $\Delta\phi$ can be expressed as
\begin{equation}
\delta\Delta\phi_{abs} = \Gamma\left(\frac{1}{\sqrt{\eta\kappa N_A}}+\frac{1}{\sqrt{\eta\kappa N_P}}\right)
\label{Eq_Collection_Efficiency_Absorption}, 
\end{equation}
where, the first term inside the parenthesis represents the quantum projection noise of the atoms, and the second term represents the shot noise of the photons (which can be regarded as the quantum projection noise of photons). The validity of this expression can be easily verified by considering various limits. Consider first the ideal case where $\varepsilon\equiv\eta\kappa=1$. For $N_P\gg N_A$, the additional noise from the laser can be neglected, and we get the fundamental noise limit due to the quantum projection noise of the atoms. On the other hand, if $N_P\ll N_A$, the quantum projection noise from the atoms can be neglected, and the process is limited by the shot-noise of the laser. In general, the parameter $\varepsilon$ represents the overall quantum efficiency of the detection process. The corresponding expression for detection via fluorescence is $\delta\Delta\phi_F=\gamma/\sqrt{\eta\rho N_A}$, where $\rho$ is the fraction of fluorescence hitting the detector.

The contribution from the second term in Eq.~(\ref{Eq_Collection_Efficiency_Absorption}) shows that the intensity of the laser beam used in absorption must be made strong enough in order to make the effect of this term negligible compared to the first term. However, since the absorption process is nonlinear and saturates for a strong laser beam, increasing the laser intensity often decreases the effective value of $\kappa$. For example, consider an ensemble of $2\times10^6$ atoms with a linear optical density of $300$, which can be realized (as we have shown above) for an ensemble confined to a cigar shaped ensemble. For a weak probe, the value of $\kappa$ is unity. However, as the probe power is increased, the value of $\kappa$ decreases dramatically. This can be seen by considering a situation where the value of $N_P$ is $10^9$, for example. Since the atomic transition used for absorption is not closed (i.e., not cyclic), the ensemble can only absorb a number of photons that is of the order of $2\times 10^6$. Thus, the maximum value of $\kappa$ would be only about $0.002$. Furthermore, if the area of the laser beam ($\pi w^2$) is much larger than the area of the atomic ensemble ($\pi\xi_T^2$), then the value of $\kappa$ can never exceed the value of $\xi_T^2/w^2$ . We are not aware of any publication reporting a cold atom interferometer that makes use of absorption for detecting the atoms, possibly because of these constraints and considerations. Nonetheless, as a matter of principle, an absorption process can certainly be used to reduce the quantum frequency fluctuation below what is observed in fluorescence detection systems, under proper choice of parameters.

\section{\label{sec:summary}Summary}
In this paper, we have described a collective state atomic interferometer (COSAIN) with N non-interacting, independent atoms in an ensemble. We have shown that the signal fringes are narrowed by $\sqrt{N}$ compared to a conventional interferometer, without entanglement. This effect is a result of the interference among collective states, and is a manifestation of interference at a Compton frequency of ten nonillion Hz, or a de Broglie wavelength of $4.5$ femtometer, for $N=10^6$ and $v = 1 m/s$.  The essence of the COSAIN is the detection of a collective state, rather than individual atomic states.  For a suitably chosen collective state, this is accomplished via a null detection scheme, wherein the detection of zero photons corresponds to the system being in this collective state.  We have presented a heterodyne detection scheme for measuring this signal.  In this scheme, the signal is detected by collecting fluorescence through stimulated Raman scattering of Stokes photons, which are emitted predominantly against the direction of the probe beam,  for a high enough resonant optical density.  We have shown that the fringe width reduction occurs due to the interference of the multiple paths among the collective states, and does not violate the fundamental quantum limit.  We have also proposed a new excitation scheme, applicable to both a conventional Raman atomic interferometer (CRAIN) as well to the COSAIN, wherein the counter-propagating beams are cross-linearly polarized.  For the COSAIN, this scheme enables an enhancement of the effective resonant optical density by placing a cavity around the atoms in the detection zone.  We have analyzed in detail the effect of various inhomegeneities, arising from the non-uniformity in experimental parameters, on the COSAIN signal, and used this analysis to identify a suitable choice of parameters for realizing a COSAIN.  The performance of the COSAIN has been compared to that of the conventional Raman atomic interferometer (CRAIN) by analyzing quantum and classical fluctuations in frequency. When the effects of detector efficiency and collection efficiency are considered, it can be seen that the COSAIN may perform $10$ times better than a CRAIN employing fluorescence detection.  

\acknowledgments{This work has been supported by the NSF grants number DGE-$0801685$ and DMR-$1121262$, and AFOSR grant number FA$9550-09-1-0652$.}


\bibliographystyle{apsrev}

\begin{thebibliography}{99}

\bibitem {Kasevich} T. L. Gustavson, P. Bouyer, and M. A.Kasevich, Phys. Rev. Lett. {\bf 78}, 2046-2049 (1997).
\bibitem {Canuel} B. Canuel et al., Phys. Rev. Lett. {\bf 97}, 010402 (2006); 
\bibitem {Chu} A. Peters, K. Y. Chung, and S. Chu, Nature 400, 849-852 (1999).
\bibitem {Snadden} M. J. Snadden et al., Phys. Rev. Lett. {\bf 81}, 971 (1998).
\bibitem {Lan} S.-Y. Lan et al., Science 339, 554-557 (2013).
\bibitem {Biraben} R. Bouchendira, P. Clad\'{e}, S. Guellati-Kh\'{e}lifa, F. Nez, and F. Biraben, Phys. Rev. Lett. 106, 080801 (2011); Ann. Phys. (Berlin) 525, 484492 (2013).
\bibitem {Cadoret}	M. Cadoret, et al. Phys. Rev. Lett. {\bf 101}, 230801 (2008).
\bibitem {Fixler} J. B. Fixler et al., Science 315, 74 (2007); G. Lamporesi, et al., Phys. Rev. Lett. {\bf 100}, 050801 (2008).
\bibitem{Muller} H. M\"{u}ller, A. Peters, and S. Chu, Nature 463, 926-929 (2010).
\bibitem{Geiger} R. Geiger et al., Nature Communications 2, 474 (2011).
\bibitem{Shahriar}	 M. S. Shahriar, M. Jheeta, Y. Tan, P. Pradhan, and A. Gangat, Opt. Comm. 243, 183 (2004).
\bibitem{KasevichChu} M. Kasevich, and S. Chu, Phys. Rev. Lett. {\bf 67}, 181-184 (1991).
\bibitem {Borde}	C.J. Bord\'{e}, Phys. Lett. A {\bf 140}, 10 (1989). 
\bibitem{Riehle} F. Riehle, Th. Kisters, A. Witte, J. Helmcke, and Ch. J. Bord\'{e}, Phys. Rev. Lett. {\bf 67}, 177 (1991).
\bibitem{Dowling}M. O. Scully, and J. P. Dowling, Phys. Rev. A {\bf 48}, 3186 (1993).
\bibitem{Malykin} G. B. Malykin, Phys. Usp. 43, 1229 (2000).
\bibitem{Broglie} L. De Broglie, Thesis, University of Paris, Paris, France (1924).
\bibitem{Eibenberger}S. Eibenberger, S. Gerlich, M. Arndt, M. Mayor, and J. T\"{u}xen, Phys. Chem. Chem. Phys. {\bf 15}, 14696 (2013).
\bibitem{Pikovski} I. Pikovski, Magdalena Zych, Fabio Costa, and \v{C}aslav Brukner, arXiv preprint arXiv:1311.1095 (2013).
\bibitem {HumeWineland}D. B. Hume, C. W. Chou, T. Rosenband, and D. J. Wineland, \pra~{\bf 80}, 052302 (2009).
\bibitem {Dicke}R. H. Dicke, Phys. Rev. 93, 99 (1954).
\bibitem {CollectiveDescription}R. Sarkar, M. E. Kim, R. Fang, Y. Tu, and S. M.Shahriar,  J. Mod. Opt. {\bf 62}, 1253 (2015).
\bibitem {Arecchi}F. T. Arecchi, E. Courtens, R. Gilmore and H. Thomas, Phys. Rev. A {\bf 6}, 2211 (1972).
\bibitem {Scully}M. O. Scully, E. S. Fry, C. H. Raymond Ooi, and K. W\'{o}dkiewicz, Phys. Rev. Lett. {\bf 96}, 010501 (2006).
\bibitem{Ueda}M. Kitagawa, and M. Ueda, \pra~{\bf 47}, 5138 (1993).
\bibitem{Polzik}J. Hald, J. L. S{\o}rensen, C. Schori, and E. S. Polzik, \prl~{\bf 83}, 1319 (1999).
\bibitem{Kuzmich}A. Kuzmich, L. Mandel, and N. P. Bigelow, \prl~{\bf 85}, 1594 (2000).
\bibitem{SpinSqueezeRev}K. Hammerer, A. S. Sorensen, and E. S. Polzik, Rev. Mod. Phys. {\bf 82}, 1041 (2010).
\bibitem{COSAC} M. E. Kim, R. Sarkar, R. Fang, and S. M. Shahriar, Phys. Rev. A. {\bf 91}, 063629 (2015).
\bibitem{CARL} R. Bonifacio, L. De Salvo, L. M. Narducci, and E. J. D'Angelo, Phys. Rev. A. {\bf 50}, 1716 (1994).
\bibitem{CARLHemmer}P. R. Hemmer, N. P. Bigelow, D. P. Katz, M. S. Shahriar, L. DeSalvo, and R. Bonifacio, \prl {\bf 77}, 1468 (1996)
\bibitem{DickePhase} K. Baumann, C. Guerlin, F Brennecke, and T. Esslinger, Nature (London) {\bf 464} (2010).
\bibitem{Olson} A. J. Olson, R. J. Niffenegger, and Y. P. Chen, Phys. Rev. A {\bf 87}, 053613 (2013).
\bibitem{Foot}C. Foot, Atomic physics (Oxford University Press, New York, 2008).
\bibitem {DLCZ}L. M. Duan, M. D. Lukin, J. I. Cirac, and P. Zoller, Nature (London) 414, 413 (2001).
\bibitem {Gardiner} C. W. Gardiner and P. Zoller, Quantum Noise (Springer-Verlag, Berlin, 1999). 
\bibitem{Wineland} W. M. Itano, J. C. Bergquist, J. J. Bollinger, J. M. Gilligan, D. J. Heinzen, F. L. Moore, M. G. Raizen, and D. J. Wineland, Phys. Rev. A {\bf 47}, 3554-3570 (1993). 





\end{thebibliography}

\end{document}